\documentclass[useAMS,usenatbib]{mn2e}
\usepackage{graphicx}
\title[Gas Kinematics and Excitation in G035.39-00.33]{Gas Kinematics and Excitation in the Filamentary IRDC G035.39-00.33}

\author[I. Jim\'enez-Serra et al.]{I. Jim\'{e}nez-Serra$^{1,2}$\thanks{E-mail: ijimenez@eso.org}, P. Caselli$^{3}$, F. Fontani$^{4}$, J. C. Tan$^{5}$, J. D. Henshaw$^{3}$, 
\newauthor J. Kainulainen$^{6}$ and A. K. Hernandez$^{7}$\\
$^{1}$European Southern Observatory, Karl-Schwarzschild-Str. 2, 85748, Garching, Germany\\
$^{2}$Harvard-Smithsonian Center for Astrophysics, 60 Garden St., 02138, Cambridge, 
MA, USA\\
$^{3}$School of Physics \& Astronomy, E.C. Stoner Building,
The University of Leeds, Leeds, LS2 9JT, UK\\
$^{4}$Osservatorio Astrofisico di Arcetri, Largo E. Fermi 5, Firenze I-50125, Italy \\
$^{5}$Department of Astronomy, University of Florida, Gainesville, FL 32611, USA\\
$^{6}$Max-Planck-Institute for Astronomy, K\"onigstuhl 17, 69117 Heidelberg, Germany \\
$^{7}$Department of Astronomy, University of Wisconsin-Madison, 475 N. Charter Street Madison, WI 53706, USA}

\begin{document}

\date{Accepted 1988 December 15. Received 1988 December 14; in original form 1988 October 11}

\pagerange{\pageref{firstpage}--\pageref{lastpage}} \pubyear{2002}

\maketitle

\label{firstpage}

\begin{abstract}

Some theories of dense molecular cloud formation involve dynamical environments driven by converging atomic flows or collisions between preexisting molecular clouds. The determination of the dynamics and physical conditions of the gas in clouds at the early stages of their evolution is essential to establish the dynamical imprints of such collisions, and to infer the processes involved in their formation. We present multi-transition $^{13}$CO and C$^{18}$O maps toward the IRDC G035.39-00.33, believed to be at the earliest stages of evolution. The $^{13}$CO and C$^{18}$O gas is distributed in three filaments (Filaments 1, 2 and 3), where the most massive cores are preferentially found at the intersecting regions between them. The filaments have a similar kinematic structure with smooth velocity gradients of $\sim$0.4-0.8$\,$km$\,$s$^{-1}$$\,$pc$^{-1}$. Several scenarios are proposed to explain these gradients, including cloud rotation, gas accretion along the filaments, global gravitational collapse, and unresolved sub-filament structures. These results are complemented by HCO$^+$, HNC, H$^{13}$CO$^+$ and HN$^{13}$C 
single-pointing data to search for gas infall signatures. The $^{13}$CO and C$^{18}$O gas motions are supersonic across G035.39-00.33, with the emission showing broader linewidths toward the edges of the IRDC. This could be due to energy dissipation at the densest regions in the cloud. The average H$_2$ densities are $\sim$5000-7000$\,$cm$^{-3}$, with Filaments 2 and 3 being denser and more massive than Filament 1. The C$^{18}$O data unveils three regions with high CO depletion factors ($f_D$$\sim$5-12), similar to those found in massive starless cores.  

\end{abstract}


\section{Introduction}

In order to explain the global properties and evolution of Giant Molecular Clouds (GMCs) and the star formation processes within these clouds, several scenarios of molecular cloud formation have been proposed. In models of flow-driven \citep[e.g.][]{hen99,bal99,heit09} and shock-induced GMC formation \citep[][]{koy00,koy02,van07}, molecular clouds are born in a highly dynamical environment characterized by the collision of large-scale, warm atomic flows that give rise to thermal and dynamical instabilities yielding filamentary molecular structures. Alternatively, \citet{tan00} and \citet{tas09} have proposed that dense molecular clumps and filaments are expected to be formed during GMC-GMC collisions, i.e. of already mostly molecular gas, induced by shear in galactic disks.

Observationally, the Herschel Space Telescope has recently revealed that the dusty interstellar medium (ISM) across the Galaxy is highly structured and organized in filaments \citep{mol10}. Among these structures, Infrared Dark Clouds (IRDCs), first detected in extinction by the Infrared Space Observatory (ISO) and by the Midcourse Space Experiment (MSX) \citep[][]{per96,egan98}, are believed to represent the initial conditions of massive star and star cluster formation. Indeed, IRDCs are cold \citep[T$\leq$25$\,$K;][]{pil07} and massive \citep[masses from some 100$\,$M$_\odot$ to 10$^5$$\,$M$_\odot$, depending on the scale of the region considered;][]{rath06,kai13}, and their clumps and cores show a range of densities \citep[from 10$^4$$\,$cm$^{-3}$ to even up to 10$^6$$\,$cm$^{-3}$; see e.g.][]{per10,rath10,but12} similar to regions known to be forming massive stars and star clusters.   

Although the general properties of IRDCs (and of their cores and clumps) have been characterized thanks to the analysis of dust extinction in the mid-IR and of dust emission at millimeter wavelengths \citep[see e.g.][]{rath06,but12,per10}, few studies have been devoted to analyze the dynamical properties of the molecular gas associated with IRDCs \citep[Hernandez \& Tan 2011; Hernandez et al. 2012;][]{kai13,per13}. 
In this context, we have recently carried out a comprehensive study of the molecular line emission and chemical composition of the gas toward the very filamentary IRDC G035.39-00.33 \citep[distance of 2.9$\,$kpc;][]{rath06}, in order to obtain a global understanding of the dynamics, history, and physical properties of the molecular gas at the initial conditions of molecular cloud formation. 

The first results obtained from this study were presented by Jim\'enez-Serra et al. (2010, Paper I), who reported the detection of faint and widespread SiO emission over parsec-scales across this IRDC. In Paper I, widespread SiO in G035.39-00.33 was interpreted as a fossil record of the large-scale shock interaction induced by a flow-flow collision that may have been involved in the formation of the IRDC\footnote{SiO is known to be an excellent tracer of shock waves in the ISM since its abundance is largely enhanced in molecular outflows \citep[even by factors $\geq$10$^6$; see e.g.][]{mar92,jim05} by the sputtering of dust grains \citep{cas97,jim08}.}. An alternative scenario considers a widespread and deeply-embedded population of low-mass stars driving molecular outflows in G035.39-00.33. This scenario cannot be ruled out at present due to the limited angular resolution of the SiO single-dish observations (Paper I).  

It is likely, however, that G035.39-00.33 is at an early stage of evolution and that its current star formation rate is relatively low. This is supported by the detection of widespread CO depletion (by up to a factor of $\sim$5) across the cloud \citep[see Hernandez et al. 2011;][]{her11}, suggesting that the molecular gas in G035.39-00.33 has been affected very little by stellar feedback. The {\it Herschel} satellite has recently mapped this IRDC at 70, 160, 250, 350 and 500$\,$$\mu$m, revealing 13 massive dense cores (masses $\geq$20$\,$M$_\odot$ and H$_2$ densities $\geq$2$\times$10$^5$$\,$cm$^{-3}$) inside the cloud \citep[][]{ngu11}. Some of these cores are indeed quiescent \citep[i.e. they contain no 24$\,$$\mu$m sources or H$_2$ shock-excited emission;][]{cha09}, and therefore are at pre-stellar/pre-cluster phase \citep[][]{rath06,but12}.

By using combined near-IR and mid-IR extinction maps, \citet{kai13} have recently obtained more accurate measurements of the total mass of G035.39-00.33. These maps have shown that the central parsec-wide region of the IRDC is close to virial equilibrium \citep[Hernandez et al. 2012;][]{her12}. The global kinematics of the molecular gas associated with this cloud are, however, rather complex, and reveals the presence of several secondary molecular filaments that could be interacting (Henshaw et al. 2013a, Paper IV). This interaction may have started $\sim$1$\,$Myr ago \citep{hen13}, implying that the gas in the IRDC would have had enough time to reach virial equilibrium with the surrounding environment \citep{her12}.

In \citet{hen13}, we concentrated on the large-scale dynamics and physical properties of the dense gas in the IRDC G035.39-00.33. In this paper (Paper V of this series), we present a multi-transition analysis of the low-density gas associated with this cloud and traced by $^{13}$CO and C$^{18}$O. These data are complemented by single-pointing spectra of typical gas infall tracers such as HCO$^+$ and HNC, and of their $^{13}$C isotopologues (H$^{13}$CO$^+$ and HN$^{13}$C) toward 
one of the most massive cores in the region (Core H6), to try to find evidence of gas infall toward this core. The observations of the $^{13}$CO and C$^{18}$O $J$=1$\rightarrow$0, $J$=2$\rightarrow$1, and $J$=3$\rightarrow$2 lines toward G035.39-0.33 are described in Section$\,$\ref{obs}. In Section$\,$\ref{res}, we present the detailed analysis of the large-scale kinematics of the $^{13}$CO and C$^{18}$O emission toward the three filaments detected in the cloud (Filaments 1, 2 and 3). In section$\,$\ref{infall}, we report the spectra of HCO$^+$, HNC, H$^{13}$CO$^+$ and HN$^{13}$C measured toward core H6 to find evidence of molecular gas infall in the core. The physical conditions of the $^{13}$CO and C$^{18}$O gas (e.g. H$_2$ number density, $^{13}$CO/C$^{18}$O column density, excitation temperature and optical depth) are presented in Section$\,$\ref{exc}. The results are discussed in Section$\,$\ref{dis}, and in Section$\,$\ref{con} we summarize our conclusions.

\section{Observations}
\label{obs}

\subsection{$^{13}$CO and C$^{18}$O line observations.}

The $J$=1$\rightarrow$0 and $J$=2$\rightarrow$1 rotational transitions of $^{13}$CO and C$^{18}$O were mapped toward the IRDC G035.39-00.33 in August and December 2008 over an area of 2$'$$\times$4$'$ (1.7$\,$pc$\times$3.4$\,$pc at a distance of 2.9$\,$kpc), with the Instituto de Radioastronom\'ia Milim\'etrica (IRAM) 30m telescope at Pico Veleta (Spain). We imaged this emission using the On-The-Fly (OTF) mode. The central coordinates of the map were $\alpha$(J2000)=18$^h$57$^m$08$^s$, $\delta$(J2000)=02$^\circ$10$'$30$''$ (l=35.517$^\circ$, b=-0.274$^\circ$), and the off-position was set at offset (1830$''$, 658$''$) with respect to the map central coordinates. While the $J$=1$\rightarrow$0 lines of $^{13}$CO and C$^{18}$O were observed with the old SIS ABCD receivers, the $J$=2$\rightarrow$1 line emission was imaged with the HERA multi-beam receiver. The receivers were tuned to single sideband (SSB) with rejections $\geq$10$\,$dB. The beam sizes were 22$''$ for the $J$=1$\rightarrow$0 lines at $\sim$110$\,$GHz, and 11$''$ for the $J$=2$\rightarrow$1 transitions at $\sim$220$\,$GHz. The VESPA spectrometer provided spectral resolutions of 20 and 80$\,$kHz, which correspond to velocity resolutions of $\sim$0.05 and 0.1$\,$km$\,$s$^{-1}$ at 110 and 220$\,$GHz, respectively. All maps were smoothed in velocity to channel widths of $\sim$0.1$\,$km$\,$s$^{-1}$. Typical system temperatures were 150-220$\,$K. 
Intensities were calibrated in units of antenna temperature, T$_A^*$, and converted into units of main beam temperature, T$_{\rm mb}$, by using beam efficiencies of 0.64 for the $^{13}$CO and C$^{18}$O $J$=1$\rightarrow$0 data, and of 0.52 for the $J$=2$\rightarrow$1 lines. 

The $J$=3$\rightarrow$2 rotational lines of $^{13}$CO and C$^{18}$O were simultaneously observed in August 2008 at the James Clerk Maxwell Telescope (JCMT) with the 16-pixel array receiver HARP-B toward the same 2$'$$\times$4$'$ area as that mapped by the IRAM 30m telescope. We imaged this emission by using the jiggle-map observing mode. Since this mode only covers a field area of 2$'$$\times$2$'$, we placed two adjacent fields centered at $\alpha_{1}$(J2000)=18$^h$57$^m$08$^s$, $\delta_{1}$(J2000)=02$^\circ$11$'$30$''$ and $\alpha_{2}$(J2000)=18$^h$57$^m$08$^s$, $\delta_{2}$(J2000)=02$^\circ$09$'$30$''$. The selected off-position was 
$\alpha$(J2000)=19$^h$02$^m$41.5$^s$ and $\delta$(J2000)=02$^\circ$57$'$38.1$''$ (i.e. l=36.85$^\circ$ and b=-1.15$^\circ$ in Galactic coordinates). The beam size of the JCMT telescope was 14$''$ at 330$\,$GHz. The spectral resolution provided by the ACSIS spectrometer was 60$\,$kHz, which corresponds to a velocity resolution of $\sim$0.05$\,$km$\,$s$^{-1}$. In order to compare these data with the $^{13}$CO and C$^{18}$O $J$=1$\rightarrow$0 and $J$=2$\rightarrow$1 maps, we smoothed the $J$=3$\rightarrow$2 line spectra to a velocity resolution of $\sim$0.1$\,$km$\,$s$^{-1}$. The system temperatures ranged from 340 to 410$\,$K. Intensities were calibrated in units of T$_A^*$, and converted into T$_{\rm mb}$ by using a beam efficiency of 0.63. A summary of the observed line frequencies, telescopes, receivers, beam sizes and beam efficiencies of the $^{13}$CO and C$^{18}$O observations, is shown in Table$\,$\ref{tab1}. 

The reduction of the $^{13}$CO and C$^{18}$O $J$=1$\rightarrow$0 and $J$=2$\rightarrow$1 IRAM 30m data was carried out with the package reduction software GILDAS\footnote{See http://www.iram.fr/IRAMFR/GILDAS}. The final molecular line data cubes were also generated with this software. The map size, pixel size, angular resolution, spectral resolution and RMS noise level (per channel and pixel) of these images are shown in Table$\,$\ref{tab2}. For the $^{13}$CO and C$^{18}$O $J$=3$\rightarrow$2 emission lines, the observed JCMT data cubes were merged and converted into fits format within the Starlink package software. These data were then reduced and used to generate the final $^{13}$CO and C$^{18}$O $J$=3$\rightarrow$2 images within GILDAS. Their data cube parameters are also reported in Table$\,$\ref{tab2}. 

\subsection{Observations of gas infall tracers.}

The molecular species HCO$^+$ and HNC are known to be good probes of infalling gas in molecular cloud cores \citep[][]{myers96,kirk13}. The infall signatures are characterized by blue-shifted asymmetries in the molecular line profiles caused by the large optical depths of the HCO$^+$ and HNC emission \citep[see e.g.][]{myers96}. In order to determine the presence of infall signatures, comparison with optically thin tracers such as H$^{13}$CO$^+$ and HN$^{13}$C, are needed. 

The $J$=1$\rightarrow$0 line emission of HCO$^+$, H$^{13}$CO$^+$ and HN$^{13}$C, and the $J$=3$\rightarrow$2 line transition of HCO$^+$, were OTF mapped with the IRAM 30m telescope toward the IRDC G035.39-00.33 in August 2008 and February 2009. The HNC $J$=1$\rightarrow$0 line emission was observed toward this cloud with the IRAM 30m in September 2013. For the HCO$^+$, H$^{13}$CO$^+$ and HN$^{13}$C line observations, we used the old SIS ABCD receivers (SSB rejections $\geq$10$\,$dB) while the HNC $J$=1$\rightarrow$0 transition was observed with the new EMIR receivers. The VESPA spectrometer provided spectral resolutions of 40$\,$kHz for the $J$=1$\rightarrow$0 lines of HCO$^+$, HNC, H$^{13}$CO$^+$ and HN$^{13}$C (velocity resolution of $\sim$0.12-0.14$\,$km$\,$s$^{-1}$), and of 80$\,$kHz for the HCO$^+$ $J$=3$\rightarrow$2 line (i.e. $\sim$0.09$\,$km$\,$s$^{-1}$). Typical system temperatures ranged from $\sim$100-140$\,$K at 90$\,$GHz, and $\sim$340-480$\,$K at 270$\,$GHz. The beam efficiencies used to convert the intensities from units of T$^*_A$ into T$_{\rm mb}$ are shown in Table$\,$\ref{tab1}. 

\begin{table}
 \centering
 \begin{minipage}{85mm}
  \caption{Molecular transitions, line frequencies, telescopes, receivers, beam sizes and beam efficiencies of our observations.}
  \begin{tabular}{lccccc}
  \hline
Transition &  $\nu$(MHz) & Telescope & Receiver & $\theta_b$('') & B$_{\rm eff}$ \\ \hline 
C$^{18}$O $J$=1$\rightarrow$0 & 109782.17 & IRAM & ABCD & 22 & 0.64 \\
$^{13}$CO $J$=1$\rightarrow$0 & 110201.35 & IRAM & ABCD & 22 & 0.64 \\
C$^{18}$O $J$=2$\rightarrow$1 & 219560.36 & IRAM & HERA & 11 & 0.52 \\
$^{13}$CO $J$=2$\rightarrow$1 & 220398.68 & IRAM & HERA & 11 & 0.52 \\
C$^{18}$O $J$=3$\rightarrow$2 & 329330.54 & JCMT & HARP-B & 14 & 0.63 \\
$^{13}$CO $J$=3$\rightarrow$2 & 330587.96 & JCMT & HARP-B & 14 & 0.63 \\ \hline

H$^{13}$CO$^+$ $J$=1$\rightarrow$0 & 86754.29 & IRAM & ABCD & 29 & 0.78 \\
HN$^{13}$C $J$=1$\rightarrow$0 & 87090.85 & IRAM & ABCD & 29 & 0.78 \\
HCO$^+$ $J$=1$\rightarrow$0 &  89188.52 & IRAM & ABCD & 28 & 0.78 \\
HNC $J$=1$\rightarrow$0 & 90663.59 & IRAM & EMIR & 27 & 0.81 \\
HCO$^+$ $J$=3$\rightarrow$2 & 267557.53 & IRAM & ABCD & 9 & 0.53 \\ \hline

\label{tab1}
\end{tabular}
\end{minipage}
\end{table}

\begin{table}
 \centering
 \begin{minipage}{85mm}
  \caption{Map size, pixel size, angular resolution, spectral resolution, and RMS noise level of the $^{13}$CO and C$^{18}$O images.}
  \begin{tabular}{rccccc}
  \hline
Transition & Map Size & Pixel Size & $\theta_b$ & $\delta v$ & RMS \\
 & ($" \times "$) & ($"$) & ($"$) & (km$\,$s$^{-1}$) & (K) \\ \hline 
$^{13}$CO (1$\rightarrow$0) & 190$\times$310 & 11 & 22 & 0.11 & 0.25  \\
          (2$\rightarrow$1) & 260$\times$380 & 5.5 & 11 & 0.11 & 0.20  \\
          (3$\rightarrow$2) & 130$\times$245 & 7 & 14 & 0.11 & 0.15  \\
C$^{18}$O (1$\rightarrow$0) & 165$\times$265 & 11 & 22 & 0.11 & 0.20 \\
          (2$\rightarrow$1) & 260$\times$375 & 5.5 & 11 & 0.11 & 0.20  \\
          (3$\rightarrow$2) & 130$\times$245 & 7 & 14 & 0.11 & 0.15  \\ \hline
\label{tab2}
\end{tabular}
\end{minipage}
\end{table}

\begin{figure}
\begin{center}
\includegraphics[angle=0,width=0.47\textwidth]{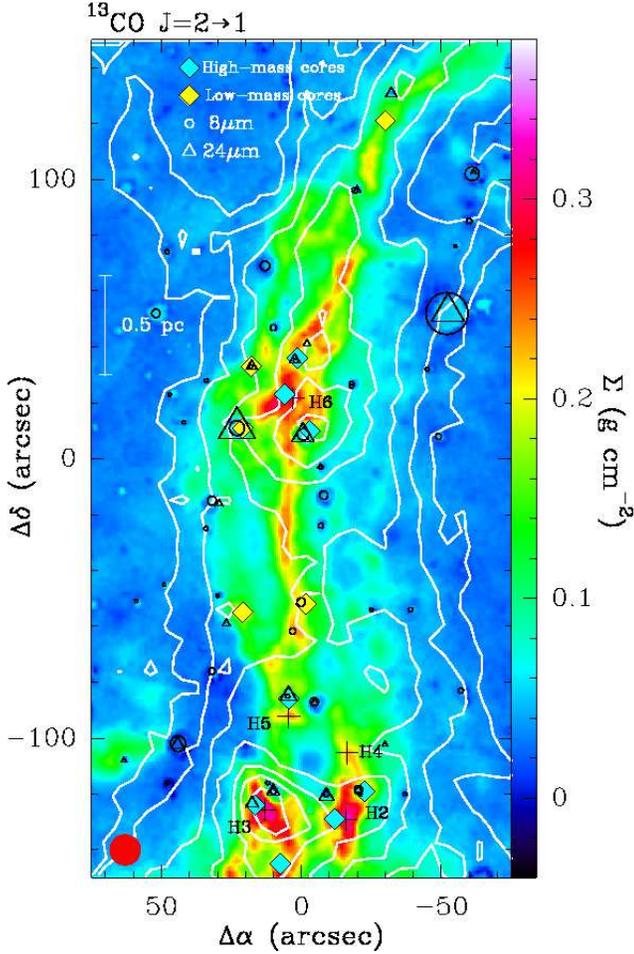}
\caption{Integrated intensity image of the $^{13}$CO $J$=2$\rightarrow$1 line emission observed toward G035.39-00.33 with the IRAM 30m telescope for the velocity range from 40 to 50$\,$km$\,$s$^{-1}$ (white contours). This image appears superimposed on the mass surface density map of the cloud \citep[in colour; angular resolution of 2$"$;][]{kai13}. For the $^{13}$CO $J$=2$\rightarrow$1 emission, the contour levels correspond to 33, 40, 50, 60, 70, 80 and 90\% the peak integrated intensity of this image (40$\,$K$\,$km$\,$s$^{-1}$). The linear intensity scale of the mass surface density map (in units of g$\,$cm$^{-2}$) is shown on the right side of the panel. Crosses indicate the location of the high-mass cores reported by \citet{but12}. Black open circles and black open triangles show the 8$\,$$\mu$m and 24$\,$$\mu$m sources detected with Spitzer toward G035.39-00.33, respectively \citep{jim10}. As in \citet{jim10}, the marker sizes have been scaled by the source flux. Yellow and light-blue filled diamonds show, respectively, the low-mass cores and the IR-quiet high-mass cores found by \citet{ngu11} with Herschel. 
The beam size of the $^{13}$CO $J$=2$\rightarrow$1 observations ($\sim$11$"$) is shown in the lower left corner of the panel.}
\label{f1}
\end{center}
\end{figure}

In this paper, we report only the single-pointing spectra measured toward the most massive core in the mapped region, core H6 \citep[][and below]{but12}, in an attempt to detect gas infall signatures in the line emission of HCO$^+$ and HNC toward this core.

\section{Results}
\label{res}

\begin{figure}
\begin{center}
\includegraphics[angle=0,width=0.48\textwidth]{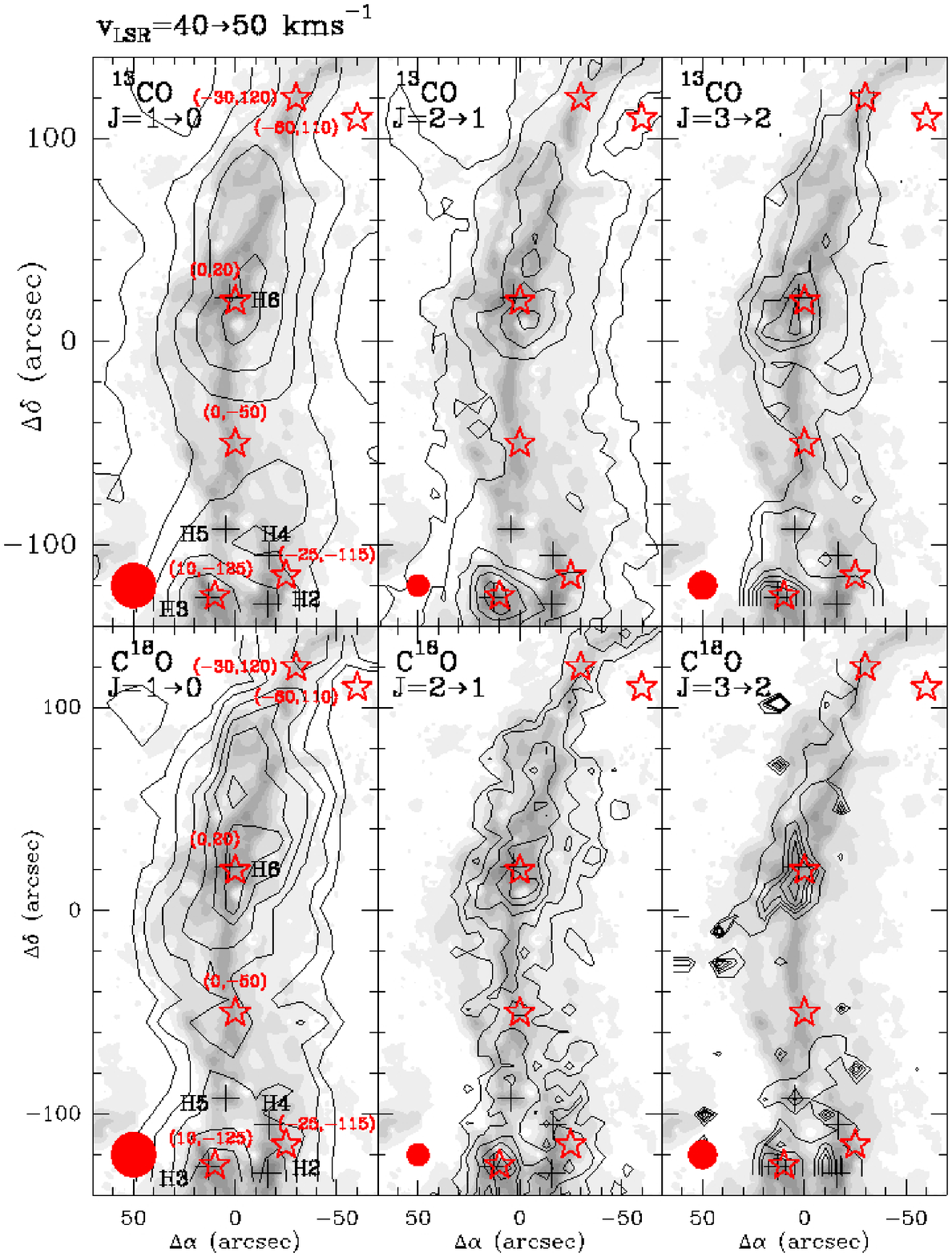}
\caption{Integrated intensity images of the $^{13}$CO and 
C$^{18}$O $J$=1$\rightarrow$0, $J$=2$\rightarrow$1 and $J$=3$\rightarrow$2 transitions for the velocity range from 40 to 50$\,$km$\,$s$^{-1}$ (black contours), superimposed on the mass surface density map of \citet[][in gray scale]{kai13}. The contour levels of the mass surface density map are 0.043, 0.074, 0.12, 0.16, 0.2, 0.25, and 0.29$\,$g$\,$cm$^{-2}$, with the first contour corresponding to an extinction of $A_v$=10$\,$mag. For the $^{13}$CO and C$^{18}$O images, the contour levels are 35, 45, 60, 70, 80 and 90\% the peak integrated intensity measured in every image, except for C$^{18}$O $J$=3$\rightarrow$2 whose contour levels correspond to 45, 60, 70, 80 and 90\% its measured peak integrated intensity. The peak integrated intensities for every image are 34.6$\,$K$\,$km$\,$s$^{-1}$ for $^{13}$CO $J$=1$\rightarrow$0, 40.0$\,$K$\,$km$\,$s$^{-1}$ for $^{13}$CO $J$=2$\rightarrow$1, 18.2$\,$K$\,$km$\,$s$^{-1}$ for $^{13}$CO $J$=3$\rightarrow$2, 6.4$\,$K$\,$km$\,$s$^{-1}$ for C$^{18}$O $J$=1$\rightarrow$0, 9.0$\,$K$\,$km$\,$s$^{-1}$ for C$^{18}$O $J$=2$\rightarrow$1, and 3.3$\,$K$\,$km$\,$s$^{-1}$ for C$^{18}$O $J$=3$\rightarrow$2. Stars show the positions selected to extract the $^{13}$CO and C$^{18}$O single-beam line spectra (Section$\,$\ref{single} and Figure$\,$\ref{f3}), and crosses indicate the location of the H cores reported by \citet{but12}. The beam sizes of the IRAM 30m and JCMT observations are shown in the lower left corner of each panel.}
\label{f2}
\end{center}
\end{figure}

In Figure$\,$\ref{f1}, we present the integrated intensity map of the $^{13}$CO $J$=2$\rightarrow$1 line emission from $v_{\rm LSR}$=40 to 50$\,$km$\,$s$^{-1}$ (white contours) observed with the IRAM 30m telescope toward the IRDC G035.39-00.33. This image has been superimposed on the mass surface density map of the cloud (in colour scale) derived by \citet{kai13} by combining a NIR extinction map with the MIR map of \citet{but12}. In Figure$\,$\ref{f1}, we also show the locations of the massive cores reported by \citet{but12} toward this cloud. The low-mass and the IR-quiet massive cores found by \citet[][see yellow and light-blue filled diamonds, respectively]{ngu11} from Herschel PACS and Spire data, are also shown. All these cores show emission at 70$\mu$m in the Herschel data. We note that the massive dense cores (filled light-blue diamonds) tend to cluster around the northern core H6 and the southern cores H2 and H3. 

From Figure$\,$\ref{f1}, we find that the observed $^{13}$CO $J$=2$\rightarrow$1 emission toward G035.39-00.33 mainly follows the filamentary structure of the IRDC seen in extinction. However, besides the main filament, two other molecular structures are detected toward the north-east and north-west of the cloud with faint counterparts seen in the mass surface density map.

The $^{13}$CO $J$=2$\rightarrow$1 main emission peaks are associated with the regions with highest extinction (i.e. densest) in the cloud at offsets (-4,10) and (10,-130). These are the regions where most of the IR-quiet massive cores are found \citep[black crosses and light-blue diamonds;][]{rath06,but12,ngu11}. While the $^{13}$CO $J$=2$\rightarrow$1 emission toward cores H3, H2 and H4 in the south of G035.39-00.33 traces the peaks of the mass surface density map, this emission is offset with respect to the densest peak in core H6. This is consistent with the significant CO depletion measured toward this core (depletion factor $f_D$$\sim$5 \citep{her11,her12}.  

In Figure$\,$\ref{f2}, we compare the morphology of the $^{13}$CO $J$=2$\rightarrow$1 emission with that from the $^{13}$CO $J$=1$\rightarrow$0 and $J$=3$\rightarrow$2 transitions, and from the C$^{18}$O $J$=1$\rightarrow$0, $J$=2$\rightarrow$1 and $J$=3$\rightarrow$2 lines. The integrated intensity emission from 40 to 50$\,$km$\,$s$^{-1}$ for all these lines is shown superimposed on the mass surface density map of \citet{kai13}. Like for $^{13}$CO $J$=2$\rightarrow$1, the other $^{13}$CO and C$^{18}$O transitions follow the filamentary structure seen in extinction. The $^{13}$CO $J$=1$\rightarrow$0 and $J$=2$\rightarrow$1 lines and the C$^{18}$O $J$=1$\rightarrow$0 transition show the molecular extensions toward the north-east and north-west of G035.39-00.33. However, the higher-J $^{13}$CO $J$=3$\rightarrow$2 and C$^{18}$O $J$=2$\rightarrow$1 and $J$=3$\rightarrow$2 lines probe denser (and therefore deeper) regions into the cloud as expected from their higher critical densities ($n_{\rm crit}$$\sim$3$\times$10$^{4}$$\,$cm$^{-2}$, i.e. a factor of $\geq$10 higher than those of the low-J lines, $n_{\rm crit}$$\sim$2$\times$10$^{3}$$\,$cm$^{-2}$). The high-J transitions of $^{13}$CO and C$^{18}$O, therefore, suffer less from contamination from the diffuse molecular envelope of the IRDC.   

\subsection{Single-beam $^{13}$CO and C$^{18}$O line spectra}
\label{single}

\begin{figure*}
\begin{center}
\includegraphics[angle=270,width=1.0\textwidth]{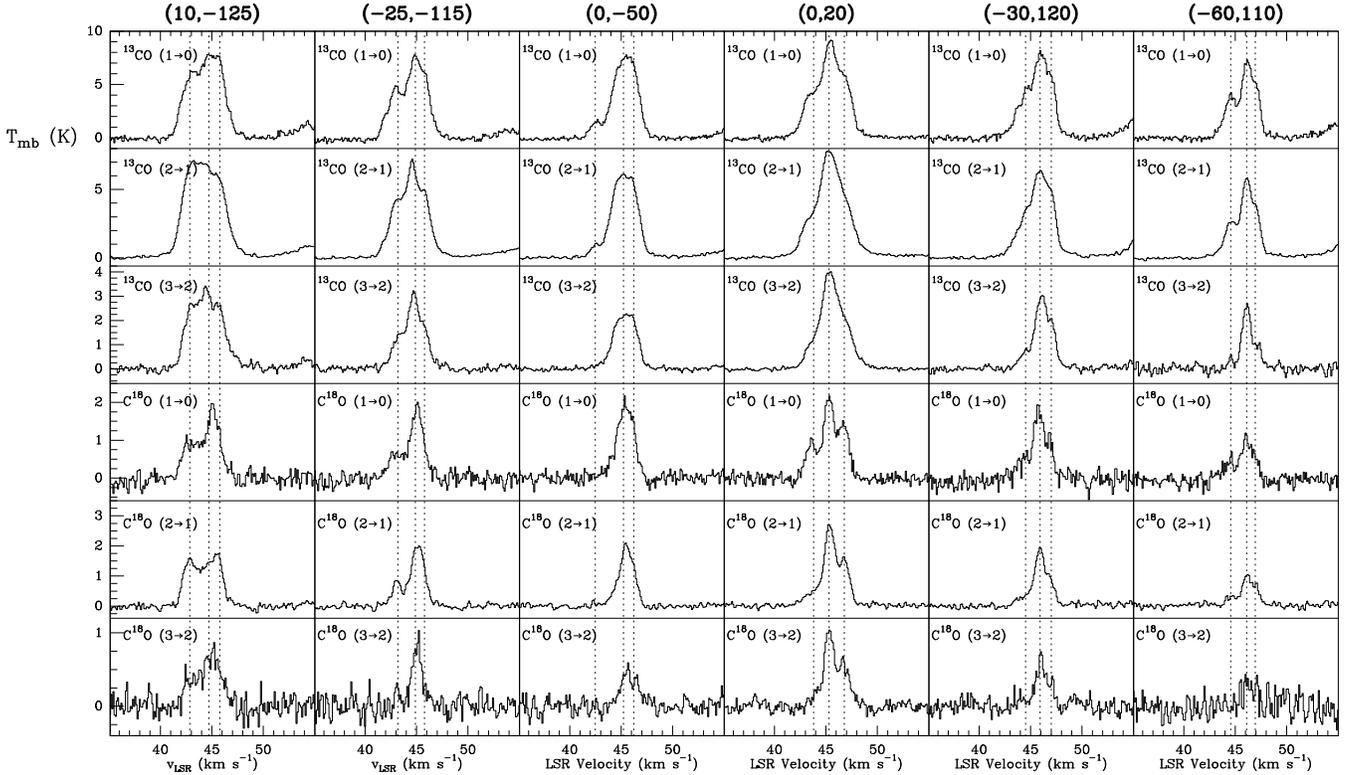}
\caption{Sample of $^{13}$CO and C$^{18}$O $J$=1$\rightarrow$0, $J$=2$\rightarrow$1 and $J$=3$\rightarrow$2 line spectra observed toward six positions across G035.39-00.33 (see stars in Figure$\,$\ref{f2}). The data have been smoothed to the 22$"$-beam of the IRAM 30m telescope at 110$\,$GHz. Vertical dotted lines show the average of the central radial velocities measured in $^{13}$CO and C$^{18}$O for every velocity component toward every position.}
\label{f3}
\end{center}
\end{figure*}

In Figure$\,$\ref{f3}, we show a sample of the $^{13}$CO and C$^{18}$O spectra observed at different positions across the IRDC. The spectra have been smoothed to the largest beam in the dataset (of 22$"$), corresponding to the $J$=1$\rightarrow$0 transition data. The positions are selected across the IRDC and are representative of cores H3 and H6 [offsets (10,-125) and (0,20)], of a position at half-distance between cores H2 and H4 [offset (-25,-115)], and of the more quiescent regions in the cloud [offsets (0,-50) and (-30,120)]. We have also selected a position off the main filament toward the northwest of the cloud [see offset (-60,110)]. 

From Figure$\,$\ref{f3}, we find that the $^{13}$CO and C$^{18}$O lines show complex molecular line profiles that change significantly across the IRDC. Consistent with the results from \citet{hen13}, three different velocity components are detected from the $^{13}$CO and C$^{18}$O emission. These velocity components are particularly clear toward (0,20), where the C$^{18}$O lines show three emission peaks. In Table$\,$\ref{tab3}, we report the central radial velocities ($v_{\rm LSR}$), linewidths ($\Delta v$), and peak intensities (T$_{\rm mb}$) of every velocity component derived by fitting the lines with three-component Gaussian profiles. The errors in $v_{\rm LSR}$ and $\Delta v$ are given by the uncertainties in the Gaussian fits, while the errors in T$_{\rm mb}$ correspond to the measured 1$\sigma$ rms level in the spectra. For the non-detections, the upper limits correspond to the 3$\sigma$ level in the $^{13}$CO and C$^{18}$O spectra. 

Table$\,$\ref{tab3} shows that the observed velocity components are centered, approximately, at radial velocities $v_{\rm LSR}$$\sim$43, 45 and 46$\,$km$\,$s$^{-1}$. However, a systematic trend is found for these velocities to become red-shifted (by $\sim$1.5$\,$km$\,$s$^{-1}$) when one moves from the south to the north of G035.39-00.33 (note the systematic shift of the vertical dotted lines in Figure$\,$\ref{f3} from left to right). For instance, for the 43$\,$km$\,$s$^{-1}$-component, the radial central velocity of $^{13}$CO $J$=1$\rightarrow$0 shifts from $v_{\rm LSR}$$\sim$42.9$\,$km$\,$s$^{-1}$ toward (10,-125) in the south to $v_{\rm LSR}$$\sim$44.5$\,$km$\,$s$^{-1}$ toward (-60,110) in the north. The same applies to the other two velocity components. As shown in Section$\,$\ref{grad}, this velocity gradient in the CO isotopologue emission is confirmed by large-scale $v_{\rm LSR}$ maps across G035.39-00.33 (velocity gradient of $\sim$0.4-0.8$\,$km$\,$s$^{-1}$$\,$pc$^{-1}$; see Section$\,$\ref{grad}). 
 
The linewidths of the $^{13}$CO and C$^{18}$O lines typically range from $\sim$1 to 2$\,$km$\,$s$^{-1}$ (see Section$\,$\ref{width} for the distribution of the $^{13}$CO and C$^{18}$O linewidths), and the brightest $^{13}$CO and C$^{18}$O emission arises from the velocity component at $\sim$45$\,$km$\,$s$^{-1}$. This component, which contains the bulk of the molecular gas, is associated with the filamentary IRDC seen in extinction (Section$\,$\ref{chan}). In the following, we describe in detail the morphology and kinematics of the different velocity components detected in $^{13}$CO and C$^{18}$O toward G035.39-00.33.

\begin{table*}
 \centering
 \begin{minipage}{200mm}
  \caption{Observed parameters of the single-beam $^{13}$CO and C$^{18}$O line spectra measured toward the selected positions shown in Figure$\,$\ref{f2}.}
  \begin{tabular}{rccccccccc}
  \hline
& \multicolumn{3}{c}{{\bf (10,-125)}} & \multicolumn{3}{c}{{\bf (-25,-115)}} & \multicolumn{3}{c}{{\bf (0,-50)}}\\ \cline{2-10}
Transition & $v_{\rm LSR}$ & $\Delta v$ & T$_{\rm mb}$ & $v_{\rm LSR}$ & $\Delta v$ & T$_{\rm mb}$ & $v_{\rm LSR}$ & $\Delta v$ & T$_{\rm mb}$ \\
 & (km$\,$s$^{-1}$) & (km$\,$s$^{-1}$) & (K) & (km$\,$s$^{-1}$) & (km$\,$s$^{-1}$) & (K) & (km$\,$s$^{-1}$) & (km$\,$s$^{-1}$) & (K) \\ \hline 

$^{13}$CO (1$\rightarrow$0) & 42.93(0.11) & 1.91(0.11) & 5.9(0.5) & 43.23(0.11) & 2.31(0.11) & 4.6(0.3) & 42.49(0.05) & 1.21(0.11) & 1.4(0.2) \\    
                            & 44.65(0.11) & 1.52(0.11) & 6.4(0.5) & 44.92(0.11) & 1.19(0.11) & 6.2(0.3) & 44.87(0.14) & 1.9(0.2)   & 6.3(0.2) \\    
                            & 45.88(0.11) & 1.49(0.11) & 5.8(0.5) & 45.97(0.11) & 1.10(0.11) & 4.6(0.3) & 46.21(0.10) & 1.5(0.10)  & 5.2(0.2) \\
$^{13}$CO (2$\rightarrow$1) & 42.61(0.10) & 1.69(0.12) & 4.7(0.3) & 43.17(0.11) & 1.89(0.11) & 4.2(0.3) & 42.48(0.09) & 1.2(0.2)   & 0.87(0.19) \\
                            & 43.9(0.3)   & 1.93(0.11) & 5.0(0.3) & 44.58(0.11) & 1.12(0.11) & 5.3(0.3) & 44.73(0.11) & 1.85(0.17) & 5.39(0.19) \\    
                            & 45.60(0.18) & 2.3(0.2)   & 5.3(0.3) & 45.81(0.11) & 1.47(0.11) & 4.5(0.3) & 46.18(0.10) & 1.57(0.11) & 4.47(0.19) \\
$^{13}$CO (3$\rightarrow$2) & 42.83(0.11) & 1.41(0.11) & 2.03(0.16) & 43.48(0.08) & 1.82(0.14) & 1.44(0.10) & $\ldots$ & $\ldots$  & $\leq$0.21 \\    
                            & 44.31(0.11) & 1.83(0.11) & 2.77(0.16) & 44.65(0.03) & 0.86(0.07) & 2.00(0.10) & 45.02(0.10) & 2.06(0.14) & 2.09(0.07) \\
                            & 45.86(0.11) & 1.83(0.11) & 1.93(0.16) & 45.56(0.12) & 1.6(0.2)   & 1.79(0.10) & 46.27(0.05) & 1.24(0.10) & 1.22(0.07) \\
C$^{18}$O (1$\rightarrow$0) & 42.95(0.09) & 1.87(0.19) & 0.97(0.16) & 42.98(0.10) & 1.7(0.2)   & 0.65(0.15) & $\ldots$ & $\ldots$  & $\leq$0.36 \\    
                            & 45.14(0.04) & 1.59(0.11) & 1.76(0.16) & 45.07(0.03) & 1.45(0.08) & 1.84(0.15) & 45.44(0.02) & 1.91(0.04) & 1.96(0.12) \\
                            & $\ldots$    & $\ldots$   & $\leq$0.48 & $\ldots$    & $\ldots$   & $\leq$0.45 & $\ldots$ & $\ldots$  & $\leq$0.36 \\
C$^{18}$O (2$\rightarrow$1) & 42.98(0.11) & 1.76(0.11) & 1.54(0.08) & 43.08(0.02) & 1.08(0.07) & 0.82(0.07) & $\ldots$ & $\ldots$  & $\leq$0.18 \\    
                            & 45.24(0.11) & 1.83(0.11) & 1.73(0.08) & 45.13(0.01) & 1.53(0.03) & 2.04(0.07) & 45.54(0.09) & 1.73(0.02) & 1.97(0.06) \\
                            & $\ldots$    & $\ldots$   & $\leq$0.24 & $\ldots$    & $\ldots$   & $\leq$0.21 & $\ldots$ & $\ldots$  & $\leq$0.18 \\
C$^{18}$O (3$\rightarrow$2) & $\ldots$    & $\ldots$   & $\leq$0.36 & $\ldots$    & $\ldots$   & $\leq$0.36 & $\ldots$ & $\ldots$  & $\leq$0.27 \\ 
                            & 45.00(0.11) & 2.2(0.2)   & 0.70(0.12) & 45.06(0.04) & 1.15(0.10) & 0.83(0.12) & 45.79(0.07) & 1.99(0.16) & 0.44(0.09) \\
                            & $\ldots$    & $\ldots$   & $\leq$0.36 & $\ldots$    & $\ldots$   & $\leq$0.36 & $\ldots$ & $\ldots$  & $\leq$0.27 \\ \hline

& \multicolumn{3}{c}{{\bf (0,20)}} & \multicolumn{3}{c}{{\bf (-30,120)}} & \multicolumn{3}{c}{{\bf (-60,110)}}\\ \cline{2-10}

$^{13}$CO (1$\rightarrow$0) & 43.67(0.03) & 2.05(0.05) & 4.23(0.15) & 44.36(0.11) & 1.96(0.11) & 3.6(0.5) & 44.49(0.11) & 1.20(0.11) & 4.0(0.4) \\    
                            & 45.37(0.02) & 1.378(0.019) & 7.82(0.15) & 46.00(0.11) & 1.60(0.11) & 6.8(0.5) & 45.95(0.11) & 1.04(0.11) & 5.7(0.4) \\    
                            & 46.83(0.03) & 1.683(0.05) & 5.49(0.15) & 47.11(0.11) & 1.17(0.11) & 3.4(0.5) & 46.88(0.11) & 1.25(0.11) & 4.9(0.4) \\    
$^{13}$CO (2$\rightarrow$1) & 43.32(0.11) & 1.89(0.11) & 2.49(0.15) & 44.42(0.11) & 2.15(0.11) & 2.7(0.3) & 44.72(0.07) & 1.78(0.16) & 2.6(0.2) \\    
                            & 45.25(0.11) & 1.86(0.11) & 6.88(0.15) & 45.91(0.11) & 1.52(0.11) & 5.3(0.3) & 46.01(0.03) & 0.84(0.04) & 4.2(0.2) \\    
                            & 46.90(0.11) & 2.28(0.11) & 3.57(0.15) & 47.07(0.11) & 1.18(0.11) & 3.6(0.3) & 46.88(0.02) & 1.28(0.06) & 3.7(0.2) \\  
$^{13}$CO (3$\rightarrow$2) & 43.95(0.11) & 2.20(0.11) & 0.89(0.05) & 44.4(0.13) & 1.3(0.2) & 0.64(0.11) & 44.57(0.11) & 0.48(0.11) & 0.52(0.15) \\    
                            & 45.33(0.11) & 1.68(0.11) & 3.19(0.05) & 46.05(0.03) & 1.35(0.12) & 3.02(0.11) & 46.03(0.11) & 0.94(0.11) & 2.43(0.15) \\
                            & 46.79(0.11) & 2.05(0.11) & 1.73(0.05) & 47.19(0.04) & 0.86(0.08) & 1.39(0.11) & 47.05(0.11) & 1.41(0.11) & 0.86(0.15) \\
C$^{18}$O (1$\rightarrow$0) & 43.59(0.04) & 1.27(0.09) & 0.95(0.10) & $\ldots$ & $\ldots$ & $\leq$0.55 & $\ldots$ & $\ldots$ & $\leq$0.42 \\    
                            & 45.29(0.02) & 1.12(0.06) & 2.14(0.10) & 45.81(0.05) & 1.14(0.14) & 1.83(0.19) & 46.19(0.05) & 1.53(0.11) & 1.00(0.14) \\
                            & 46.74(0.03) & 1.19(0.07) & 1.45(0.10) & 46.94(0.06) & 0.66(0.12) & 0.93(0.19) & $\ldots$ & $\ldots$ & $\leq$0.42\\
C$^{18}$O (2$\rightarrow$1) & 44.2(0.4)   & 2.7(0.5)   & 0.53(0.06) & 44.9(0.3) & 2.3(0.3) & 0.41(0.07) & 44.42(0.07) & 0.86(0.11) & 0.29(0.08)  \\
                            & 45.39(0.01) & 1.07(0.05) & 2.39(0.06) & 45.82(0.02) & 0.87(0.05) & 1.38(0.07) & 46.33(0.03) & 1.74(0.07) &  0.96(0.08) \\
                            & 46.80(0.02) & 1.22(0.04) & 1.51(0.06) & 46.73(0.02) & 1.37(0.06) & 0.91(0.07) & $\ldots$ & $\ldots$ & $\leq$0.24 \\
C$^{18}$O (3$\rightarrow$2) & 44.06(0.11) & 1.0(0.3)   & 0.19(0.06) & $\ldots$ & $\ldots$ & $\leq$0.24 & $\ldots$ & $\ldots$ & $\leq$0.42 \\
                            & 45.28(0.03) & 0.94(0.07) & 1.01(0.06) & 45.94(0.11) & 0.96(0.11) & 0.62(0.08) & $\ldots$ & $\ldots$ & $\leq$0.42 \\
                            & 46.68(0.06) & 1.54(0.15) & 0.57(0.06) & 46.96(0.11) & 0.90(0.11) & 0.29(0.08) & $\ldots$ & $\ldots$ & $\leq$0.42  \\
\hline
\label{tab3}
\end{tabular}
\end{minipage}
\end{table*}

\subsection{Channel maps of the $^{13}$CO and C$^{18}$O emission}
\label{chan}

\begin{figure*}
\begin{center}
\includegraphics[angle=0,width=0.5\textwidth]{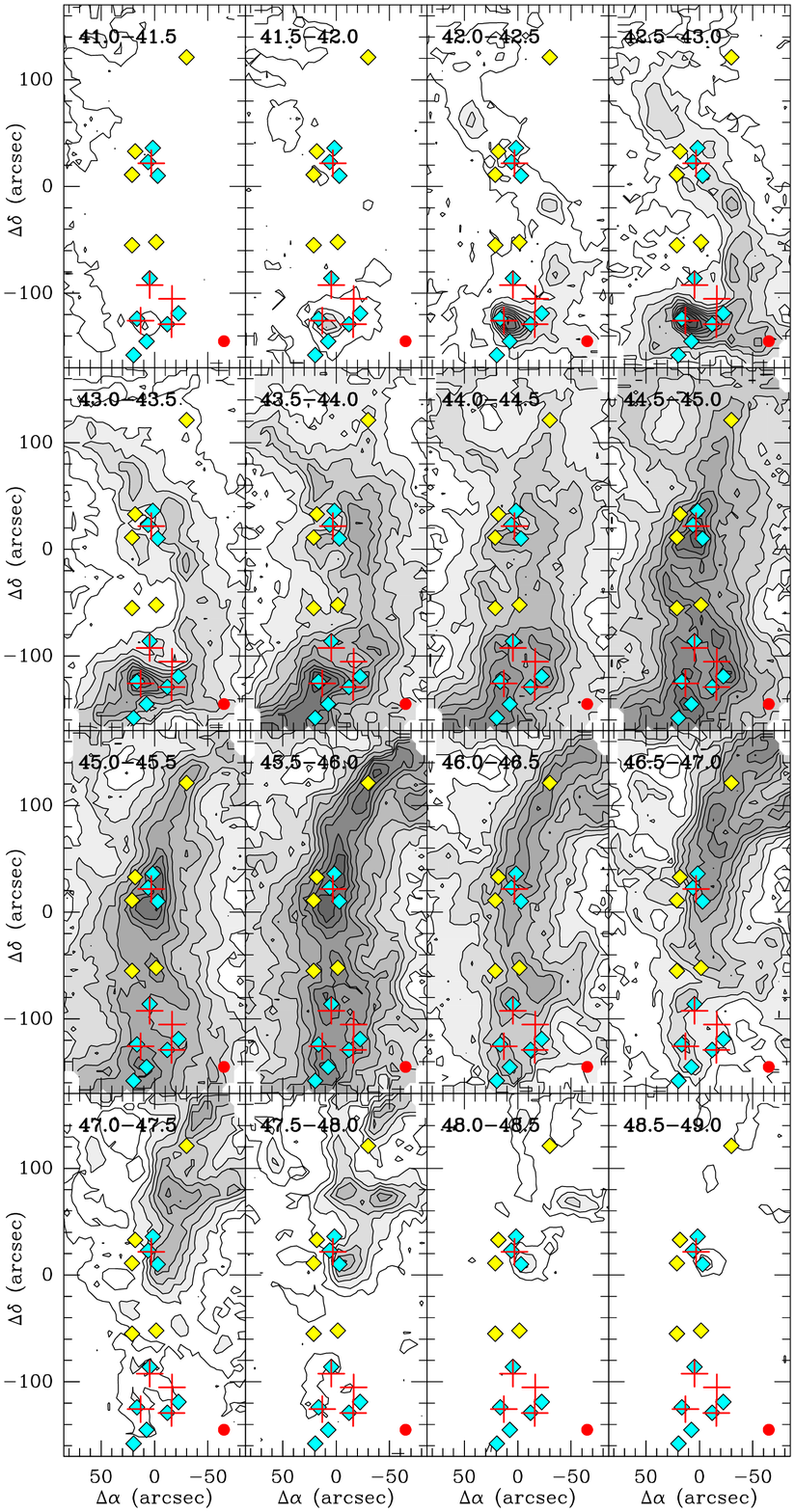}
\caption{Channel maps of the $^{13}$CO $J$=2$\rightarrow$1 line emission toward G035.39-00.33 measured between 41$\,$km$\,$s$^{-1}$ and 49$\,$km$\,$s$^{-1}$ in velocity increments of 0.5$\,$km$\,$s$^{-1}$. The velocity range of every map is shown at the upper left corner of every panel. The first contour and step levels are 0.50 (5$\sigma$) and 0.5$\,$K$\,$km$\,$s$^{-1}$, respectively. Red crosses indicate the position of the dense cores reported by \citet{but12}, and yellow filled diamonds and light-blue filled diamonds show the location of the low-mass and massive dense cores detected with {\it Herschel} \citep{ngu11}. The beam size of the $^{13}$CO $J$=2$\rightarrow$1 observations ($\sim$11$''$) is shown at the lower right corner in every panel.}
\label{f4}
\end{center}
\end{figure*}

\begin{figure*}
\begin{center}
\includegraphics[angle=0,width=0.5\textwidth]{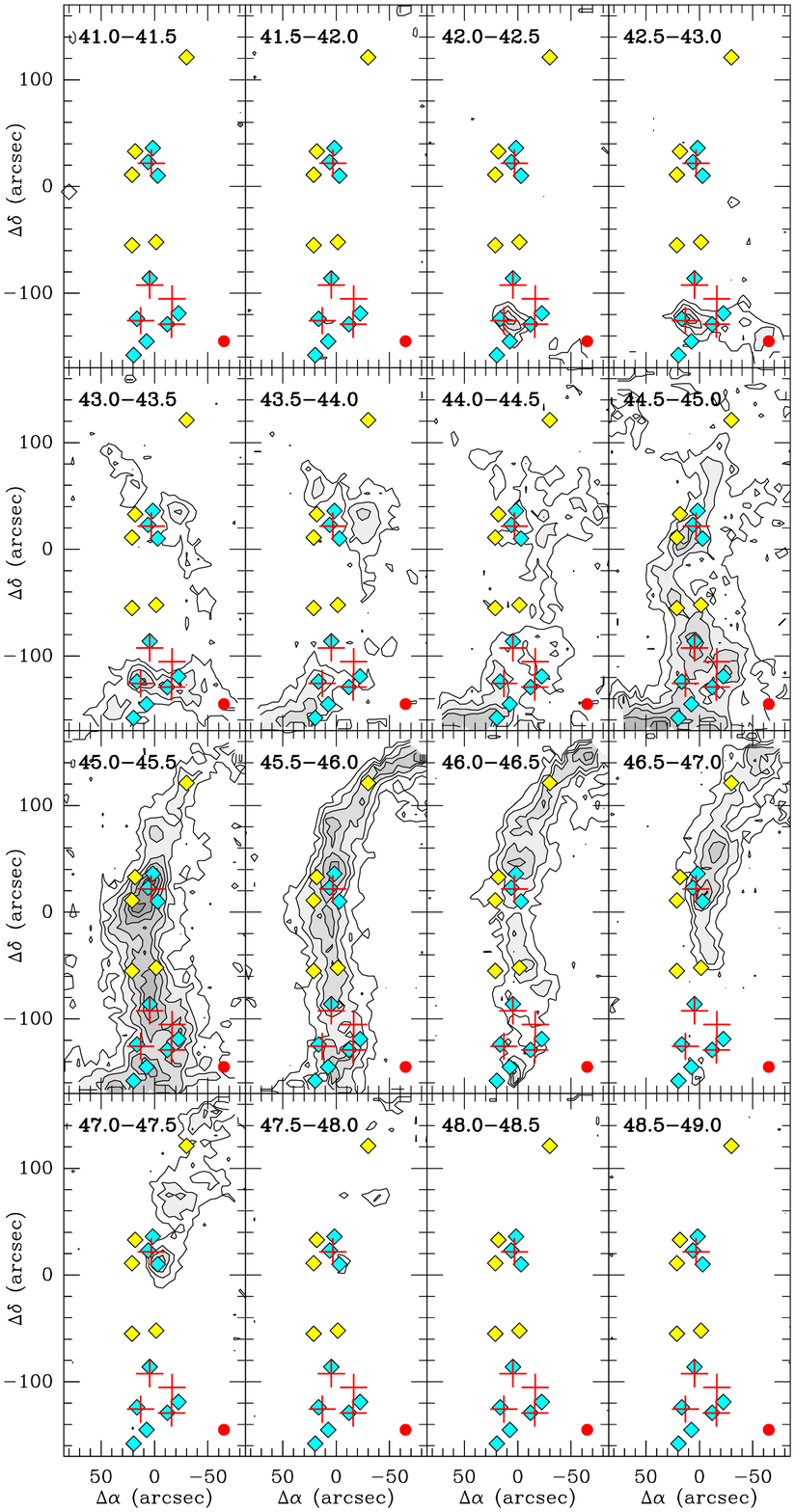}
\caption{As in Figure$\,$\ref{f4} but for the C$^{18}$O $J$=2$\rightarrow$1 line emission. The first contour and step levels are 0.40 (4$\sigma$) and 0.3$\,$K$\,$km$\,$s$^{-1}$, respectively. Crosses and symbols are as in Figure$\,$\ref{f4}.}
\label{f5}
\end{center}
\end{figure*}

In Figures$\,$\ref{f4} and \ref{f5}, we show the channel maps of the $^{13}$CO and C$^{18}$O $J$=2$\rightarrow$1 line emission measured toward G035.39-00.33 between 41 and 49$\,$km$\,$s$^{-1}$ in velocity increments of 0.5$\,$km$\,$s$^{-1}$. We use the $^{13}$CO and C$^{18}$O $J$=2$\rightarrow$1 line data cubes because they have the highest angular resolution available (beam of $\sim$11$"$) within our CO isotopologue molecular line dataset. 

\begin{figure}
\begin{center}
\includegraphics[angle=0,width=0.48\textwidth]{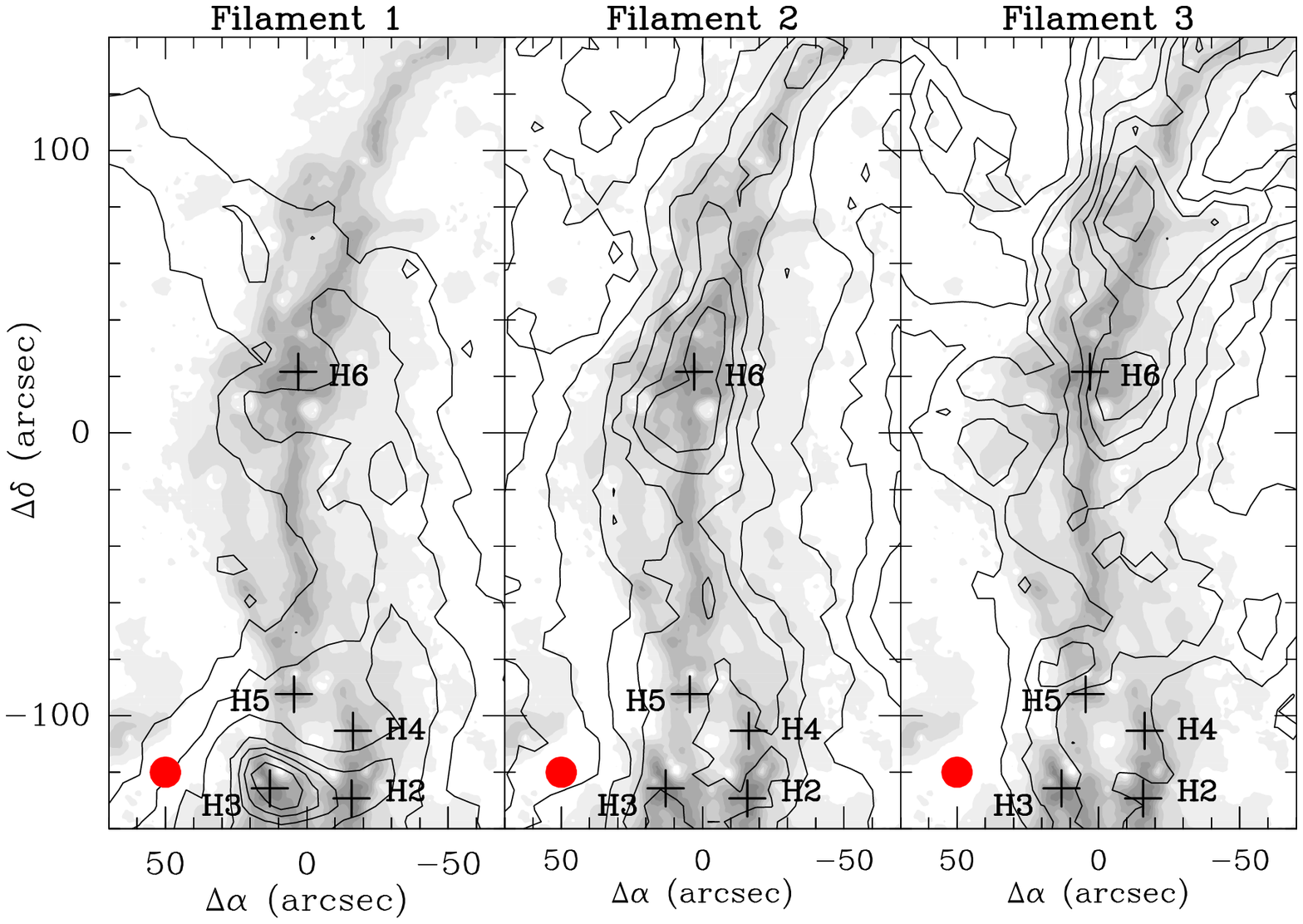}
\caption{Integrated intensity images of the molecular Filaments 1, 2 and 3 seen in $^{13}$CO $J$=2$\rightarrow$1 toward G035.39-00.33. The velocity ranges of the maps are 42.5-44.5$\,$km$\,$s$^{-1}$ for Filament 1, 44.5-46.0$\,$km$\,$s$^{-1}$ for Filament 2, and 46.0-48.0$\,$km$\,$s$^{-1}$ for Filament 3. The mass surface density map of \citet{kai13} is shown in gray scale (same as in Figure$\,$\ref{f2}). The contour levels of the $^{}$CO $J$=2$\rightarrow$1 emission are at 25, 35, 45, 60, 70, 80, and 90\% the peak integrated intensity in every map (i.e. 20.5$\,$K$\,$km$\,$s$^{-1}$, 12.6$\,$K$\,$km$\,$s$^{-1}$, and 11.4$\,$K$\,$km$\,$s$^{-1}$ for Filaments 1, 2 and 3 respectively). Crosses indicate the locations of the massive cores reported by \citet{but12}. The beam size of the IRAM 30m observations ($\sim$11$"$) is shown in the lower left corner in every  panel.}
\label{f6}
\end{center}
\end{figure}

As noted in \citet{hen13}, the general kinematics of the CO emission are characterized by the presence of several molecular filaments associated with the IRDC (Figure$\,$\ref{f4}). The CO gas concentrates along, at least, two clear elongated/filamentary structures extending over $\sim$300$"$ (i.e. $\sim$4.4$\,$pc) along the north-south direction. The first filament is seen at the velocity component at $v_{\rm LSR}$=43$\,$km$\,$s$^{-1}$ (Filament 1). This filament runs from the north-east to the south-east, curving to the west towards the centre of the filament and showing a completely different morphology to that seen for the IRDC in extinction (the IRDC in extinction curves toward the east; see Figure$\,$\ref{f1}). At more red-shifted velocities ($v_{\rm LSR}$$\sim$44.0-45.0$\,$km$\,$s$^{-1}$), the CO molecular gas evolves from this filamentary morphology into a more complex structure. The dominant feature at these velocities has the same morphology as the IRDC seen in extinction and therefore it is associated with the densest part of the cloud \citep[Filament 2; see][]{hen13}. Core H6 is approximately located at the intersecting region between Filaments 1 and 2 (Figure$\,$\ref{f4}). For velocities $v_{\rm LSR}$$\geq$46$\,$km$\,$s$^{-1}$ (Filament 3), the main filament develops two secondary extensions/arms toward the north-west and south-west of core H6. Although this new structure largely overlaps with Filament 2, it likely represents an additional feature different from Filaments 1 and 2 since it appears as an independent structure in velocity space with respect to the other two filaments \citep[][]{hen13}. Figure$\,$\ref{f6} shows the integrated intensity images of the three individual molecular filaments (Filaments 1, 2 and 3) identified in G035.39-00.33 from the $^{13}$CO $J$=2$\rightarrow$1 emission.

From Figures$\,$\ref{f4} and \ref{f5}, it is clear that the IR-quiet massive dense cores detected toward G035.39-00.33 tend to be found toward the regions where the filaments intersect, and where most of the molecular gas is concentrated. In particular, the chain of cores found toward the south follows the merged $^{13}$CO structure for Filaments 1 and 2 (see Figure$\,$\ref{f4} for the velocity range 44.5-45.0$\,$km$\,$s$^{-1}$), suggesting that these structures are physically connected and likely interacting. The detection of broader $^{13}$CO and C$^{18}$O line profiles toward these regions [offsets (10,-125) and (0,20) in Figure$\,$\ref{f3}], could be due to the merging of the filaments, although on-going star-formation could also be responsible for the broadening of the lines. As reported in \citet{jim10}, broad SiO emission has been detected toward core H6 (SiO condensation named N) and toward core H4 (condensation S). 

\subsection{Distribution of v$_{\rm LSR}$ across the CO filaments}
\label{grad}

By using the single-dish spectra of $^{13}$CO and C$^{18}$O measured across G035.39-00.33, we can derive the distribution of the central radial velocity, $v_{\rm LSR}$, and linewidth, $\Delta v$, of the molecular gas for every position in this IRDC. We use the $J$=2$\rightarrow$1 and $J$=3$\rightarrow$2 transitions of $^{13}$CO and C$^{18}$O because they probe denser gas than the $J$=1$\rightarrow$0 lines, and their maps have a higher angular resolution (beam of $\sim$11$"$-14$"$; Table$\,$\ref{tab1}). In this Section, we will report the results from the $v_{\rm LSR}$ maps, while the distribution of $\Delta v$ for the $^{13}$CO and C$^{18}$O lines will be presented in Section$\,$\ref{width}.

To calculate the $v_{\rm LSR}$ and $\Delta v$ maps across G035.39-00.33, we have smoothed the $^{13}$CO and C$^{18}$O single-dish $J$=2$\rightarrow$1 and $J$=3$\rightarrow$2 data to the same angular resolution of 14$"$ (i.e. to the largest beam of the JCMT maps), so that the $J$=2$\rightarrow$1 and $J$=3$\rightarrow$2 spectra can be compared directly. The individual spectra are simultaneously fitted with three-component Gaussian line profiles as described in Appendix$\,$\ref{fits}. In our analysis of the $^{13}$CO lines, we have considered line spectra with peak intensities $\geq$9$\sigma$ only (with $\sigma$ the rms in every spectrum). This criterion was first used in \citet{hen13} and takes into account that the typical detection threshold for Gaussian line profiles is 3$\sigma$ and that the $^{13}$CO emission toward G035.39-00.33 has three overlapping velocity components. For C$^{18}$O, the applied detection threshold is $\geq$5$\sigma$ since the C$^{18}$O lines are weaker (Appendix$\,$\ref{fits}). The derived values of $v_{\rm LSR}$ and $\Delta v$ are then used to generate the $v_{\rm LSR}$ and $\Delta v$ maps. This Gaussian decomposition allows to study not only the individual kinematic structure of every filament, but also the excitation conditions of the CO gas within them (see Section$\,$\ref{exc}). 

\begin{figure*}
\begin{center}
\includegraphics[angle=0,width=0.7\textwidth]{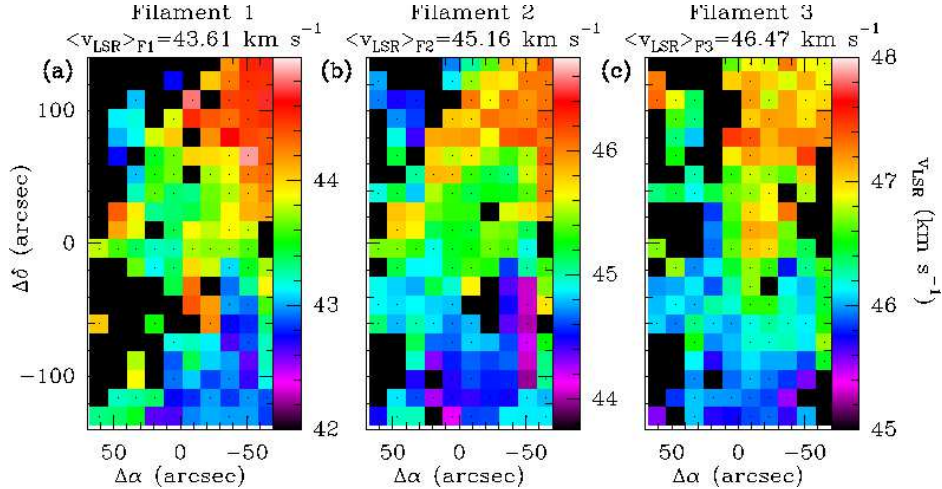}
\caption{Spatial distribution of $v_{\rm LSR}$ derived from $^{13}$CO $J$=2$\rightarrow$1 for Filaments 1, 2 and 3 (panels {\bf (a)}, {\bf (b)}, and {\bf (c)}, respectively). Only data with intensities $\geq$9$\sigma$ are plotted. Color scale is shown on the right of every panel. Note that the velocity range varies from panel to panel. The intensity-weighted average of the radial velocity for every filament, $<v_{\rm LSR}>_{\rm F_k}$, is shown at the upper part of panels {\bf (a)}, {\bf (b)}, and {\bf (c)} (Table$\,$\ref{tab4}). The kinematics of the molecular gas reveal a smooth north-south velocity gradient along the IRDC, consistent with gas accretion onto core H6 along the filaments (see Section$\,$\ref{grad}).}
\label{f7}
\end{center}
\end{figure*}

In Figure$\,$\ref{f7}, we show the distribution of $v_{\rm LSR}$ derived from the $^{13}$CO $J$=2$\rightarrow$1 data for Filaments 1, 2 and 3. The intensity-weighted average of $v_{\rm LSR}$ for every filament is shown in the upper part of every panel, and is calculated as described in Appendix$\,$\ref{aver}. The intensity-weighted average of $v_{\rm LSR}$ is 43.61$\,$km$\,$s$^{-1}$ for Filament 1, 45.16$\,$km$\,$s$^{-1}$ for Filament 2, and 46.47$\,$km$\,$s$^{-1}$ for Filament 3 (Table$\,$\ref{tab4}). In agreement with the results reported in \citet{hen13}, the filaments are separated in velocity space by $\sim$3$\,$km$\,$s$^{-1}$. We note that these values change by $<$0.2\% if data with peak intensities $\geq$5$\sigma$, instead of $\geq$9$\sigma$, are considered in our analysis.

For all three filaments, our data reveal a velocity gradient in the north-south direction with the red-shifted gas peaking toward the north-northwest of G035.39-00.33, and with the blue-shifted emission toward the south of this cloud. While the most red- and blue-shifted velocities are measured at distances further away from core H6, the $^{13}$CO gas motions in the vicinity of this core show radial velocities close to the average values (i.e. the gas radial velocities get more red- or blue-shifted with increasing distance to core H6; see Figure$\,$\ref{f7}). Although the velocity gradient is relatively smooth for Filament 2, we note that the kinematics of the $^{13}$CO gas in the vicinity of this core for Filaments 1 and 3 are more complex. This could be due either to feedback from local star formation \citep[broad SiO emission is detected toward this position; see][]{jim10} or to smaller-scale structures unresolved in our single-dish observations (see Section$\,$\ref{subfil} and Henshaw et al. 2013b).

The derived values for the velocity gradient are $\sim$0.4-0.8$\,$km$\,$s$^{-1}$$\,$pc$^{-1}$ from the northern part of G035.39-00.33 to core H6, and $\sim$0.6-0.8$\,$km$\,$s$^{-1}$$\,$pc$^{-1}$ from the southern region to this core. This velocity gradient is consistent with that previously reported in \citet[][]{hen13} (of $\sim$1$\,$km$\,$s$^{-1}$$\,$pc$^{-1}$) for Filament 2. Similar velocity gradients have also been reported toward the Orion molecular cloud \citep[$\sim$0.7$\,$km$\,$s$^{-1}$$\,$pc$^{-1}$;][]{bally87}, the DR21 filament \citep[$\sim$0.8-2.3$\,$km$\,$s$^{-1}$$\,$pc$^{-1}$;][]{sch10}, or even toward the low-mass Serpens South cluster-forming region \citep[$\sim$1.4$\,$km$\,$s$^{-1}$$\,$pc$^{-1}$;][]{kirk13}. This smooth velocity gradient in G035.39-00.33 has become apparent thanks to the new $^{13}$CO and C$^{18}$O data, and contrasts with the sharp velocity transition found in \citet[][]{hen13}. This is likely due to the lower-angular resolution of the N$_2$H$^{+}$ and C$^{18}$O $J$=1$\rightarrow$0 data compared to our $^{13}$CO $J$=2$\rightarrow$1 line data ($\sim$26$"$ vs. 14$"$, respectively). We also stress that the smooth velocity gradient reported for every filament cannot be attributed to an artifact of our three-component Gaussian fit method, since the close inspection of the single-pointing spectra (Section$\,$\ref{single}) already revealed a systematic trend for the radial velocities of the filaments to shift from blue- to red-shifted velocities as one moves from the southern to the northern regions in the IRDC.  

Besides $^{13}$CO $J$=2$\rightarrow$1, the analysis of the $^{13}$CO $J$=3$\rightarrow$2 and C$^{18}$O $J$=2$\rightarrow$1 data also reveals a similar behavior for the individual motions of the gas within every filament (see Figures$\,$\ref{a2} and \ref{a3}). This demonstrates that our analysis of the kinematics of the molecular gas toward G035.39-00.33 does not depend on the transition used, and provides robust results against the multi-Gaussian profile fitting method described in Appendix$\,$\ref{fits}. Indeed, the average $v_{\rm LSR}$ derived for every filament is very similar in all transitions and CO isotopologues (see Table$\,$\ref{tab4}), except for Filament 2 where the C$^{18}$O $J$=2$\rightarrow$1 emission appears globally red-shifted (by $\sim$0.2$\,$km$\,$s$^{-1}$) compared to the lower-density $^{13}$CO $J$=2$\rightarrow$1 line. This behavior is similar to that reported between C$^{18}$O $J$=1$\rightarrow$0 and N$_2$H$^{+}$ $J$=1$\rightarrow$0, and has been interpreted as a signature of the ongoing interaction between Filaments 1 and 2 \citep{hen13}. This velocity shift cannot be attributed to large optical depth effects since $^{13}$CO and C$^{18}$O are moderately optically thick and optically thin, respectively (see Sections$\,$\ref{13co} and \ref{c18o}).

The emission from the C$^{18}$O $J$=3$\rightarrow$2 is very faint throughout the map except for few positions (see Figure$\,$\ref{f2}). This has prevented us from generating the individual motions of the three filaments from this transition. 

\subsection{Distribution of $\Delta v$ in the CO filaments}
\label{width}

\begin{table*}
 \centering
 \begin{minipage}{110mm}
  \caption{Intensity-weighted average of $v_{\rm LSR}$, $\Delta v$, and $\sigma_{\rm NT}$ for Filaments 1, 2 and 3.}
  \begin{tabular}{rccccccccc}
  \hline
& \multicolumn{3}{c}{$<v_{\rm LSR}>$} & \multicolumn{3}{c}{$<\Delta v>_{\rm F_k}$} & \multicolumn{3}{c}{$\sigma_{\rm NT}$} \\
& \multicolumn{3}{c}{(km$\,$s$^{-1}$)} & \multicolumn{3}{c}{(km$\,$s$^{-1}$)} & \multicolumn{3}{c}{(km$\,$s$^{-1}$)} \\
Transition & F1 & F2 & F3 & F1 & F2 & F3 & F1 & F2 & F3 \\ \hline
$^{13}$CO (2$\rightarrow$1) & 43.61 & 45.16 & 46.47 & 1.78 & 1.75 & 1.59 & 0.71 & 0.70 & 0.62 \\
$^{13}$CO (3$\rightarrow$2) & 43.72 & 45.24 & 46.54 & 1.47 & 1.48 & 1.40 & 0.57 & 0.57 & 0.54 \\
C$^{18}$O (2$\rightarrow$1) & 43.71 & 45.38 & 46.42 & 1.30 & 1.21 & 1.15 & 0.49 & 0.44 & 0.42 \\ \hline
\label{tab4}
\end{tabular}
Note.- $<v_{\rm LSR}>$, $<\Delta v>$, and $\sigma_{\rm NT}$ are calculated for positions where the peak intensities are $\geq$9$\sigma$ for $^{13}$CO $J$=2$\rightarrow$1 and $J$=3$\rightarrow$2, and $\geq$5$\sigma$ for C$^{18}$O $J$=2$\rightarrow$1. 
\end{minipage}
\end{table*}

We have generated maps of the average linewidth of the $^{13}$CO and C$^{18}$O lines, $<\Delta v>_{\rm Tot}$, by using the linewidths measured for every velocity component (or filament) and position in the map (see Appendix$\,$\ref{aver} for the method to calculate this average). By doing this, we can infer the global level of turbulence across the cloud, and look for any systematic changes in the molecular linewidths between the densest regions in the IRDC and its lower-density envelope. 

\begin{figure*}
\begin{center}
\includegraphics[angle=0,width=0.7\textwidth]{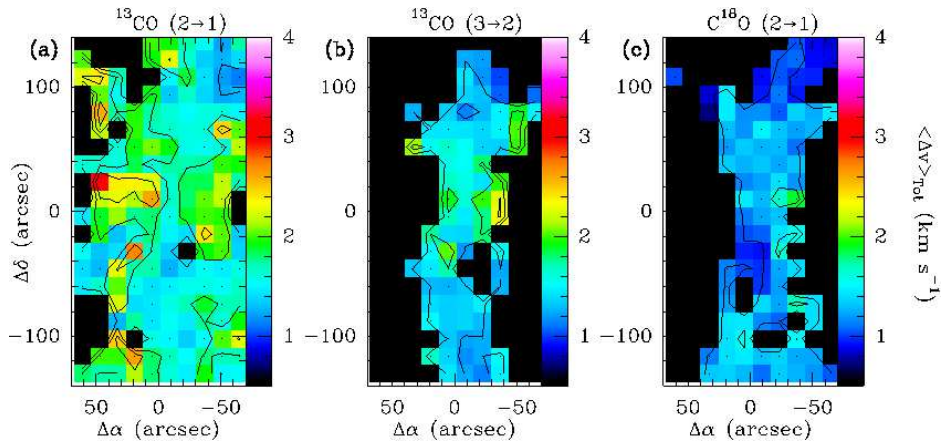}
\caption{Maps of the intensity-weighted average of the linewidth (in km$\,$s$^{-1}$) obtained across the IRDC G035.39-00.33, $<\Delta v>_{\rm Tot}$ (in colour) for the $^{13}$CO $J$=2$\rightarrow$1 (panel {\bf (a)}), $^{13}$CO $J$=3$\rightarrow$2 (panel {\bf (b)}), and C$^{18}$O $J$=2$\rightarrow$1 lines (panel {\bf (c)}). Colour scale is shown on the right in every panel. Contours are 1.25,  1.75, and 2.25$\,$km$\,$s$^{-1}$ for $^{13}$CO $J$=2$\rightarrow$1 and $^{13}$CO $J$=3$\rightarrow$2, with the narrower linewidths found toward the central regions of the IRDC. For C$^{18}$O $J$=2$\rightarrow$1, contours are 1.0, 1.5, and 2.0$\,$km$\,$s$^{-1}$, showing a flatter linewidth distribution.} 
\label{f8}
\end{center}
\end{figure*}

In Figure$\,$\ref{f8}, we report the maps of $<\Delta v>_{\rm Tot}$ derived for $^{13}$CO $J$=2$\rightarrow$1, $^{13}$CO $J$=3$\rightarrow$2, and C$^{18}$O $J$=2$\rightarrow$1 toward G035.39-00.33. From Figure$\,$\ref{f8}, we find that the general distribution of $<\Delta v>_{\rm Tot}$ for $^{13}$CO $J$=2$\rightarrow$1 is not uniform and shows broader line emission along the outer edges of the IRDC. This behaviour was already noted in Paper II, and resembles that reported by \citet{pin10} toward the B5 low-mass star forming core. Since we only consider the positions in the $^{13}$CO $J$=2$\rightarrow$1 map with line detections $\geq$9$\sigma$, the line broadening toward the outer regions in G035.39-00.33 is not due to poorer quality of the Gaussian fits, but to a real increase in the velocity dispersion in the lower density gas. The spatial distribution of $<\Delta v>_{\rm Tot}$ for $^{13}$CO $J$=3$\rightarrow$2 and C$^{18}$O $J$=2$\rightarrow$1 shows narrower linewidths toward the central regions of G035.39-00.33, supporting this idea. Indeed, the average linewidths $<\Delta v>_{\rm F_k}$ of the individual filaments in G035.39-00.33 (see Appendix$\,$\ref{aver} and Table$\,$\ref{tab4}) are significantly narrower for $^{13}$CO $J$=3$\rightarrow$2 and C$^{18}$O $J$=2$\rightarrow$1 (i.e. $\sim$1.4-1.5 and 1.2-1.3$\,$km$\,$s$^{-1}$, respectively) than for $^{13}$CO $J$=2$\rightarrow$1 ($\sim$1.6-1.8$\,$km$\,$s$^{-1}$), which could be related to energy dissipation. The derived values of $<\Delta v>_{\rm F_k}$ are found to vary by $<$3\% if data with peak intensities $\geq$5$\sigma$, instead of $\geq$9$\sigma$, are considered in our analysis.

The trend of increasing linewidth toward the low-density outer regions of the IRDC can also be seen from the distribution of the non-thermal (turbulent) velocity dispersion, $\sigma_{\rm NT}$, for the individual Filaments 1, 2 and 3. $\sigma_{\rm NT}$ can be determined from the observed linewidth, $\Delta v$, as follows \citep[][]{myers83}:

   \begin{equation}
      \sigma_{\rm NT}=\sqrt{\frac{\Delta v^2}{8\,\,ln2} - \frac{k_B\,\,T_{\rm kin}}{m}},
   \label{Dave}
   \end{equation}
 
\noindent 
where $k_B$ is the Boltzmann's constant, $T_{\rm kin}$ the kinetic temperature of the molecular gas, and $m$ the mass of the molecular species (either 29$\,$a.m.u. for $^{13}$CO or 30$\,$a.m.u. for C$^{18}$O). The thermal dispersion of the molecular gas, $\sigma_T$=$\sqrt{k_BT_{\rm kin}/m}$, is $\sim$0.07$\,$km$\,$s$^{-1}$ assuming that the gas kinetic temperature is 15$\,$K \citep{pil06,rag11,fon12}. Table$\,$\ref{tab4} reports the values of $\sigma_{\rm NT}$ derived for every filament.

\begin{figure*}
\begin{center}
\includegraphics[angle=0,width=0.75\textwidth]{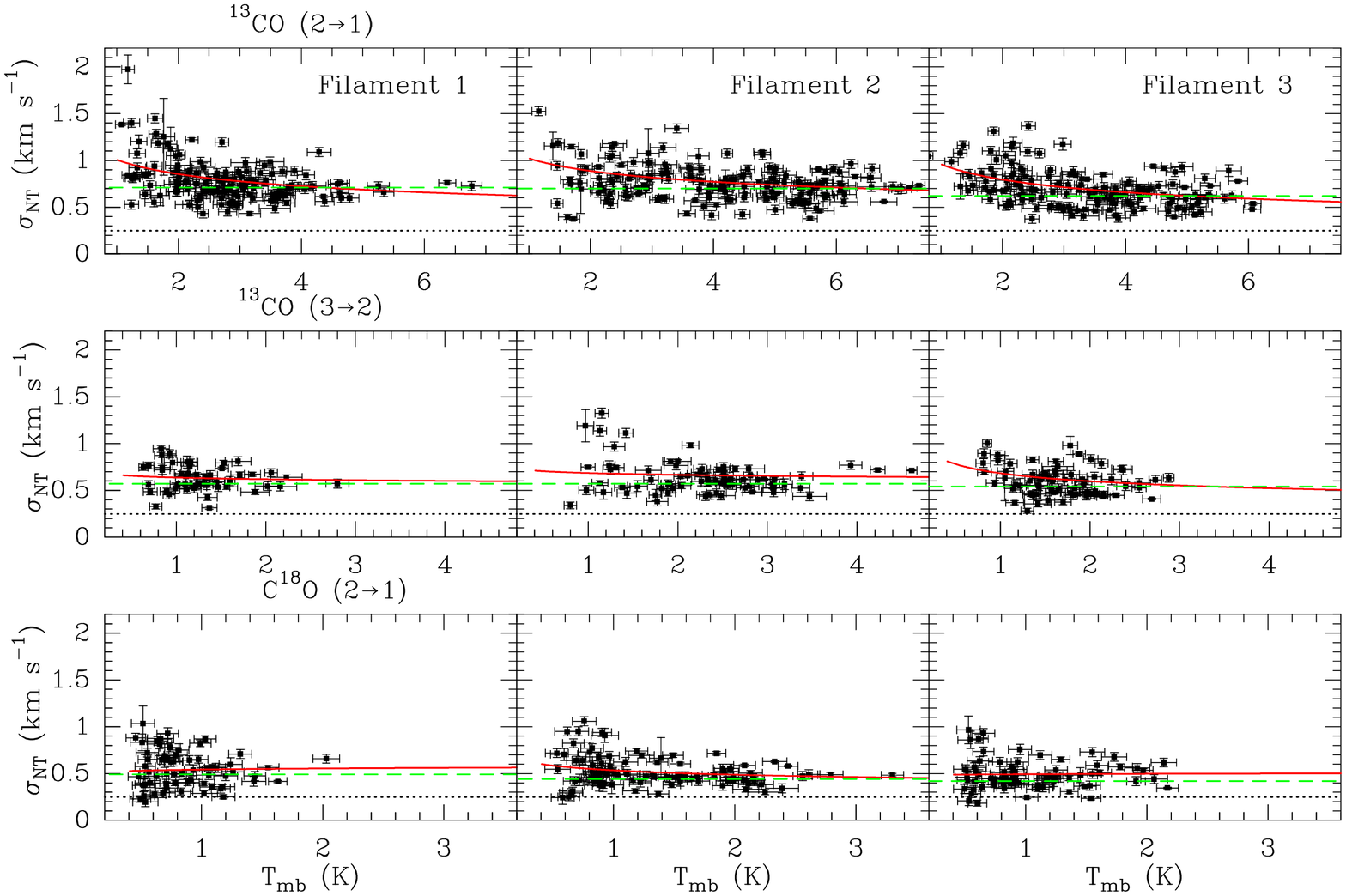}
\caption{Non-thermal velocity dispersion derived from $^{13}$CO $J$=2$\rightarrow$1 and $J$=3$\rightarrow$2, and from C$^{18}$O $J$=2$\rightarrow$1, toward every position in Filaments 1, 2 and 3, and represented as a function of peak intensity (in units of T$_{\rm mb}$). For $^{13}$CO $J$=2$\rightarrow$1 and $J$=3$\rightarrow$2, we only consider data points with T$_{\rm mb}$$\geq$9$\sigma$, while for C$^{18}$O $J$=2$\rightarrow$1, we impose that T$_{\rm mb}$$\geq$5$\sigma$. Horizontal dotted lines show the value of the sound speed of the average particle ($\mu$=2.33) at T$_{\rm kin}$=15$\,$K ($c_{s}$=0.25$\,$km$\,$s$^{-1}$). Green dashed lines show the average non-thermal dispersion of every filament and transition (Table$\,$\ref{tab4}). Red lines show the power law fit to the data in the form $\sigma_{\rm NT}$=$a \times (T_{\rm mb})^b$ (see text for {\it a} and {\it b} values).}
\label{f9}
\end{center}
\end{figure*}

In Figure$\,$\ref{f9}, we show the non-thermal velocity dispersion, $\sigma_{\rm NT}$, derived from the $^{13}$CO and C$^{18}$O data as a function of peak intensity, T$_{\rm mb}$, for every filament. To our knowledge, this is the first attempt made to measure the non-thermal velocity dispersion across different filaments within the same IRDC. The peak intensity, T$_{\rm mb}$, of these lines is used as a proxy of column density and, therefore, of the radial distance to the axis of the filaments. In order to better show the increasing trend of $\sigma_{\rm NT}$ for decreasing T$_{\rm mb}$, we have carried out a power law fit to the data in the form $\sigma_{\rm NT}$=$a \times (T_{\rm mb})^b$ (see red lines in Figure$\,$\ref{f9}). For the fits, we have binned the data in temperature ranges of 0.25$\,$K. While the average $b$ parameter for $^{13}$CO $J$=2$\rightarrow$1 is $\sim$$-$0.24, for $^{13}$CO $J$=3$\rightarrow$2 this parameter is $b$$\sim$$-$0.09, indicating a weaker dependency of $\sigma_{\rm NT}$ with radial distance. As expected for a higher-density tracer, C$^{18}$O $J$=2$\rightarrow$1 shows almost a flat distribution with an average $b$ value $\sim$$-$0.03.     

From Figure$\,$\ref{f9}, we find that the gas in the individual molecular filaments of G035.39-00.33 is supersonic, with an average non-thermal velocity dispersion $\sigma_{\rm NT}$ (see also Table$\,$\ref{tab4}) being factors $\sim$2-3 larger than the sound speed in the medium (i.e. sound Mach number $\sim$2-3 with the sound speed being $c_s$=0.25$\,$km$\,$s$^{-1}$)\footnote{The sound speed, $c_s$, is calculated assuming an average particle mass $\mu$=2.33 and a kinetic temperature T$_{\rm kin}$=15$\,$K.}. This is in agreement with the results by \citet{vas11} and \citet{rag12} toward two samples of IRDCs, and contrasts with the gas motions toward the molecular filaments in the L1517 and the L1495/B213 low-mass star forming regions, where these motions are, respectively, subsonic and mildly transonic toward both the high-density cores (seen in N$_2$H$^+$) and the lower-density {\it envelope} of the filaments probed by C$^{18}$O \citep[see][]{hac11,hac13}. 

\subsection{Gas infall tracer lines toward core H6}
\label{infall}

In Section$\,$\ref{grad}, we have reported the detection of a relatively smooth velocity gradient of $\sim$0.4-0.8$\,$km$\,$s$^{-1}$$\,$pc$^{-1}$ in the north-south direction, common to Filaments 1, 2 and 3. As discussed in Section$\,$\ref{acc}, one scenario that could explain this velocity gradient is gas accretion flows along Filaments 1, 2 and 3 onto the massive core H6. We investigate whether typical gas infall tracers such as HCO$^+$ and HNC indeed show blue-shifted asymmetries in their self-absorbed molecular line profiles as a result of gas accretion onto this core.

\begin{figure}
\begin{center}
\includegraphics[angle=0,width=0.4\textwidth]{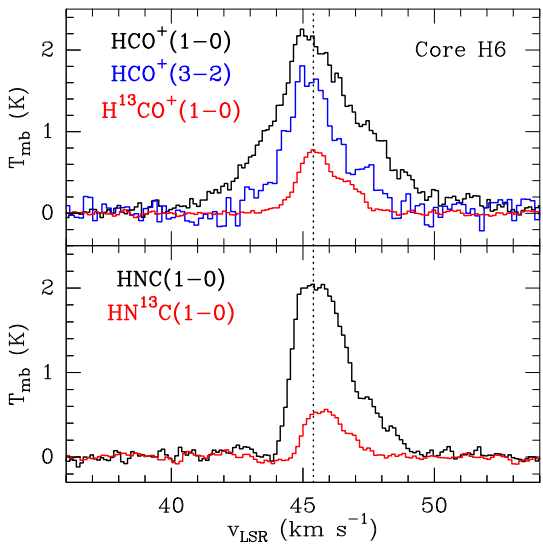}
\caption{Spectra of the HCO$^+$ $J$=1$\rightarrow$0, HCO$^+$ $J$=3$\rightarrow$2, H$^{13}$CO$^+$ $J$=1$\rightarrow$0, HNC $J$=1$\rightarrow$0, and HN$^{13}$C $J$=1$\rightarrow$0 lines measured toward core H6. The spectra have been smoothed to the largest beam of $\sim$29$"$ of the IRAM 30m data. The vertical dotted line indicates the $v_{\rm LSR}$ of the dense C$^{18}$O $J$=2$\rightarrow$1 gas in Filament 2 ($v_{\rm LSR}$$\sim$45.4$\,$km$\,$s$^{-1}$; see Table$\,$\ref{tab4}).}
\label{fig-infall}
\end{center}
\end{figure}

In Figure$\,$\ref{fig-infall} we report the spectra of the $J$=1$\rightarrow$0 lines of HCO$^{+}$, HNC, H$^{13}$CO$^{+}$ and HN$^{13}$C, and of the $J$=3$\rightarrow$2 transition of HCO$^{+}$, measured toward core H6, and smoothed to the same angular resolution of 29$"$ (the largest beam of the IRAM 30m observations). All species show similar line profiles. In particular, the emission from optically thick tracers, HCO$^+$ and HNC, peaks at the same $v_{\rm LSR}$ ($\sim$45.4$\,$km$\,$s$^{-1}$, i.e. the $v_{\rm LSR}$ derived for the dense C$^{18}$O $J$=2$\rightarrow$1 gas; see Table$\,$\ref{tab4}) as their optically thin $^{13}$C isotopologues, H$^{13}$CO$^+$ and HN$^{13}$C. This clearly contrasts with the behaviour seen for self-absorbed spectra toward contracting cloud cores, for which the optically thin species show their emission peaks red-shifted compared to the optically thick lines \citep[see][]{myers95,kirk13,per13}. The asymmetry in the line profiles of HCO$^+$ and HNC found toward core H6 are probably not due to gas infalling onto this core (at least at angular scales of $\geq$29$"$, or $\geq$0.4$\,$pc), but to the intrinsic kinematic structure of the IRDC (Section$\,$\ref{chan}). We note that infall motions are not likely to appear at the denser regions in this IRDC since the HCO$^+$ $J$=3$\rightarrow$2 emission does not show any self-absorption toward core H6 either. 

\section{Excitation conditions of the CO gas in G035.39-00.33}
\label{exc}

We can now use the derived peak intensities and linewidths of the $^{13}$CO and C$^{18}$O $J$=2$\rightarrow$1 and $J$=3$\rightarrow$2 lines, to estimate the physical conditions of the gas such as H$_2$ number density, n(H$_2$), $^{13}$CO/C$^{18}$O column density, N($^{13}$CO) or N(C$^{18}$O), excitation temperature, T$_{\rm ex}$, and optical depth, $\tau$, in the three molecular filaments detected toward G035.39-00.33 (Sections$\,$\ref{single} and \ref{chan}). To do this, we have used the Large Velocity Gradient (LVG) approximation \citep{sob47}, for which we have considered the collisional coefficients of $^{13}$CO and C$^{18}$O with H$_2$ calculated by \citet{yan10}, and the first 18 rotational levels of these two molecules. We assume a kinetic temperature of the gas, T$_{\rm kin}$, of $\sim$15$\,$K, which is similar to those found in other IRDCs from NH$_3$ observations \citep{pil06,rag11,fon12}, and which is in close equilibrium with the temperature of the dust measured toward G035.39-00.33 \citep[T$_{d}$$\sim$16-17$\,$K; see][]{ngu11}. In Section$\,$\ref{tkin}, we however evaluate the sensitivity of our LVG results to small changes in the kinetic temperature.  

The average linewidth between the $J$=2$\rightarrow$1 and $J$=3$\rightarrow$2 lines of $^{13}$CO and C$^{18}$O was used as an input parameter in the LVG code. The $J$=1$\rightarrow$0 line emission from $^{13}$CO and C$^{18}$O was not employed in our calculations because 1) this transition probes lower-density gas in the IRDC \citep{her11}; and 2) all data would have to be smoothed to the poorer angular resolution of $\sim$22$"$ of the $J$=1$\rightarrow$0 observations (Table$\,$\ref{tab1}). In Section$\,$\ref{J10}, we examine in what extent our LVG results are modified by the inclusion of the $J$=1$\rightarrow$0 data in the LVG analysis.

A grid of LVG models were run with H$_2$ number densities ranging from 900$\,$cm$^{-3}$ to $\sim$10$^5$$\,$cm$^{-3}$, and with $^{13}$CO and C$^{18}$O column densities ranging from 10$^{10}$$\,$cm$^{-2}$ to $\sim$10$^{20}$$\,$cm$^{-2}$. For every grid position, the LVG code calculates the peak intensity of the rotational lines of a molecular species (i.e. $^{13}$CO or C$^{18}$O in our case) with quantum numbers from $J_{\rm up}$=1 to $J_{\rm up}$=18. The LVG solution is then obtained by means of a minimization technique by comparing the predicted intensities of the $J$=2$\rightarrow$1 line and the predicted line intensity ratio T(3$\rightarrow$2)/T(2$\rightarrow$1), with those observed toward every position in the map. For $^{13}$CO, only those positions across the IRDC where both transitions were detected above the 9$\sigma$ level, were considered in the calculations (i.e. we use the same dataset as that reported in Sections$\,$\ref{grad} and \ref{width}). Note that this is important because it guarantees that the derived peak intensities and linewidths of the lines, key for the determination of the physical conditions of the gas, have been measured with good accuracy. The LVG results provide maps of n(H$_2$), N($^{13}$CO), T$_{\rm ex}$ and $\tau$ (see Section$\,$\ref{13co}). However, in the case of C$^{18}$O, the threshold of 5$\sigma$ for both transitions was exceeded only toward three positions across the IRDC (see Section$\,$\ref{c18o}), including 1) core H6; 2) the C$^{18}$O emission peak toward the west of core H6; and 3) 10$"$ north the position where narrow SiO emission lines have been detected \citep[][]{jim10}. 

\subsection{LVG results for $^{13}$CO}
\label{13co}

\begin{table*}
 \centering
 \begin{minipage}{150mm}
  \caption{Average physical properties of Filaments 1, 2 and 3 derived from $^{13}$CO for T$_{\rm kin}$=13$\,$K, 15$\,$K and 17$\,$K.}
  \begin{tabular}{ccccccccccc}
  \hline
Filament & T$_{\rm kin}$ & n(H$_2$) & N($^{13}$CO) & T$_{\rm ex}$(2$\rightarrow$1) & T$_{\rm ex}$(3$\rightarrow$2) & $\tau_{\rm 21}$ & $\tau_{\rm 32}$ & $\Sigma_{\rm ^{13}CO}$ & Mass \\
& (K) & (cm$^{-3}$) & (cm$^{-2}$) & (K) & (K) & & & (g$\,$cm$^{-2}$) & (M$_\odot$) \\ \hline

1 & 13 & 8000 & 1.1$\times$10$^{16}$ & 9.7 & 7.9 & 2.2 & 1.4 & 0.014 & 5700 \\
2 & 13 & 8600 & 2.4$\times$10$^{16}$ & 10.4 & 8.5 & 4.5 & 3.3 & 0.030 & 27000 \\
3 & 13 & 11500 & 1.2$\times$10$^{16}$ & 10.2 & 8.4 & 2.7 & 1.8 & 0.015 & 13300 \\ \hline

1 & 15 & 5100 & 9.4$\times$10$^{15}$ & 9.7 & 7.8 & 2.0 & 1.3 & 0.012 & 4300 \\
2 & 15 & 7500 & 1.4$\times$10$^{16}$ & 11.0 & 8.9 & 2.7 & 2.0 & 0.020 & 15000 \\
3 & 15 & 7300 & 9.6$\times$10$^{15}$ & 10.7 & 8.6 & 2.2 & 1.5 & 0.012 & 10800 \\ \hline

1 & 17 & 4000 & 8.2$\times$10$^{15}$ & 9.9 & 7.9 & 1.8 & 1.1 & 0.010 & 3800 \\
2 & 17 & 5600 & 1.3$\times$10$^{16}$ & 11.3 & 9.0 & 2.5 & 2.0 & 0.016 & 14500 \\
3 & 17 & 5500 & 8.9$\times$10$^{15}$ & 10.9 & 8.7 & 2.0 & 1.4 & 0.011 & 9900 \\ \hline

\label{tab5}
\end{tabular}
\end{minipage}
\end{table*}

\begin{figure}
\begin{center}
\includegraphics[angle=0,width=0.5\textwidth]{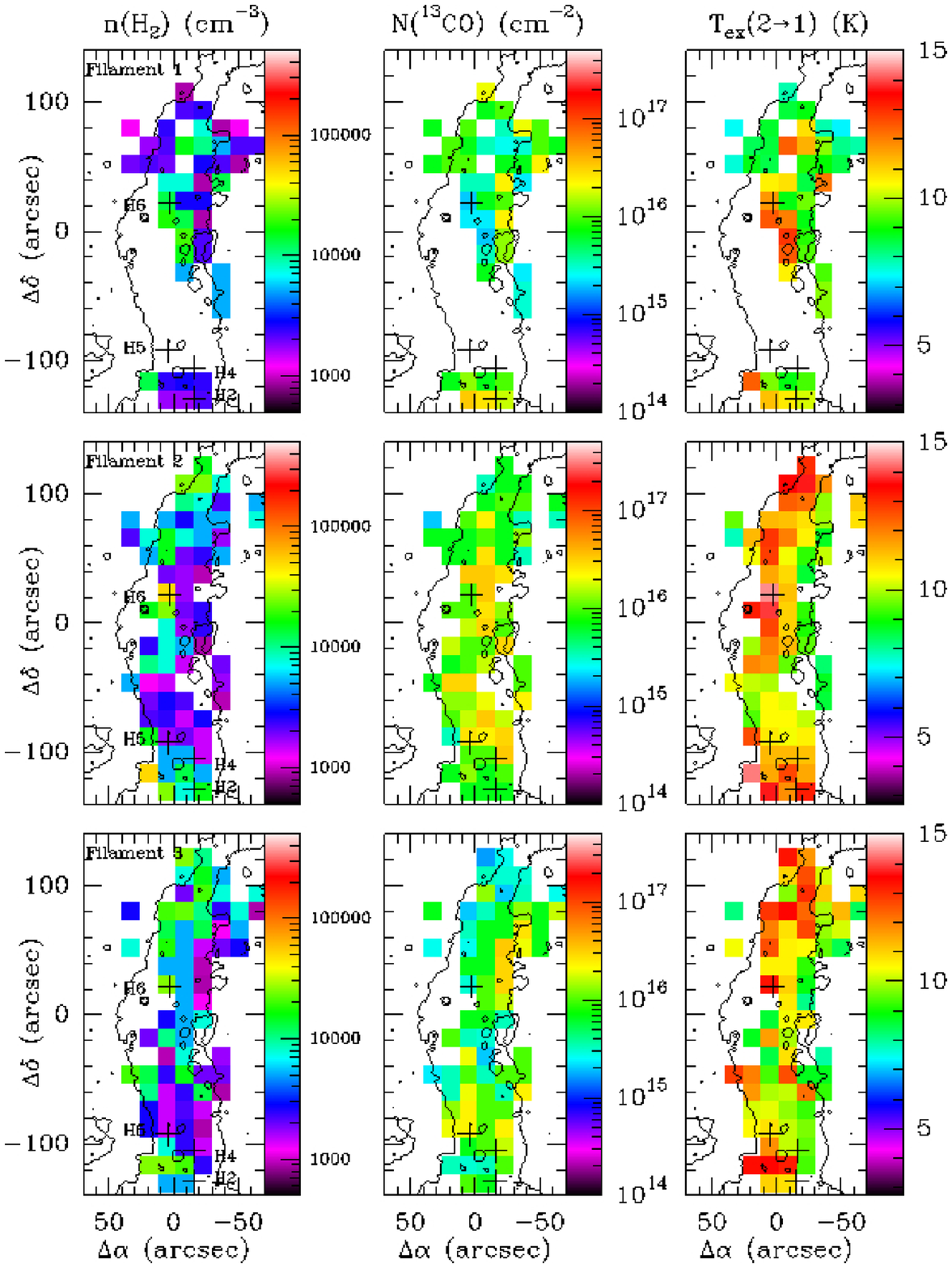}
\caption{H$_2$ number density, n(H$_2$), $^{13}$CO column density, N($^{13}$CO), and excitation temperature, T$_{\rm ex}$, derived from $^{13}$CO $J$=2$\rightarrow$1 and $J$=3$\rightarrow$2 for Filament 1 (upper panels), Filament 2 (middle panels) and Filament 3 (lower panels) using the LVG approximation. Overlapped onto these maps, we also show the mass surface density map derived by \citet{kai13} in silhouette. Colour scale is shown on the right side of every panel. Crosses indicate the location of the cores reported by \citet{but12}.}
\label{f10}
\end{center}
\end{figure}

In Figure$\,$\ref{f10}, we present the H$_2$ number density, n(H$_2$), $^{13}$CO column density, N($^{13}$CO), and excitation temperature of the gas for the $J$=2$\rightarrow$1 transition, T$_{\rm ex}$(2$\rightarrow$1), obtained for Filament 1 (upper panels), Filament 2 (middle panels), and Filament 3 (lower panels) by using the LVG approximation. Table$\,$\ref{tab5} also reports the average values of these parameters measured for every filament. From Figure$\,$\ref{f10}, we find that the derived n(H$_2$) typically range between some 10$^3$$\,$cm$^{-3}$ to few 10$^4$$\,$cm$^{-3}$, as expected for a low-density tracer such as $^{13}$CO, and in agreement with those H$_2$ densities estimated toward low-mass molecular clouds such as Perseus \citep[see e.g.][]{ben06,pin08}. The average H$_2$ density for Filament 1 is somewhat smaller ($\sim$5100$\,$cm$^{-3}$) than those of Filaments 2 and 3 ($\sim$7300-7500$\,$cm$^{-3}$; see Table$\,$\ref{tab5}), indicating that the former filament is less dense than the latter ones. This was already noted in \citet{hen13} from the lack of N$_2$H$^+$ $J$=1$\rightarrow$0 emission in Filament 1. 

The derived $^{13}$CO column densities range from $\sim$2-3$\times$10$^{15}$$\,$cm$^{-2}$ to $\sim$2-3$\times$10$^{16}$$\,$cm$^{-2}$ in the three filaments. The average values of N($^{13}$CO) are very similar in the three filaments and are $\sim$10$^{16}$$\,$cm$^{-2}$ (Table$\,$\ref{tab5}). Assuming an isotopic ratio $^{12}$C/$^{13}$C of 53 \citep{wil94}, a CO abundance of $\chi$(CO)$\sim$2$\times$10$^{-4}$ \citep{lac94}, and a mass per H nucleus of $\mu_H$=2.34$\times$10$^{-24}$$\,$g, we can derive the average mass surface density for every filament as:

   \begin{equation}
      \Sigma_{\rm 13CO}=1.24\times10^{-2}\times\left[\frac{N(\rm ^{13}CO)}{10^{16} (\rm cm^{-2})}\right]\,\,\rm g\,\,\,cm^{-2}.
   \label{surf}
   \end{equation}
 
\noindent
The average values of N($^{13}$CO) imply mass surface densities of $\sim$0.012$\,$g$\,$cm$^{-2}$ for Filaments 1 and 3, and of $\sim$0.02$\,$g$\,$cm$^{-2}$ for Filament 2 (Table$\,$\ref{tab5}). By adding up all these contributions, we obtain a total surface mass density $\sim$0.04$\,$g$\,$cm$^{-2}$, which is consistent with the average value obtained in \citet[][]{her11} from their C$^{18}$O data.

An estimate of the {\it relative} H$_2$ gas masses between the filaments can be derived from the values of N($^{13}$CO) shown in Figure$\,$\ref{f10}. As shown in Table$\,$\ref{tab5}, the derived H$_2$ gas masses for Filaments 1, 2 and 3 are, respectively, 4300, 15000 and 10800$\,$M$_\odot$. This implies that Filament 1 is a factor of $\sim$2-3 less massive than Filaments 2 and 3, while the latter two filaments have very similar gas masses. One may think that the lower mass for Filament 1 could be a consequence of the smaller number of data points considered in the calculation. However, by using a similar area than that covered by Filaments 2 and 3, and by assuming an average column density\footnote{This $^{13}$CO column density has been derived by using the LVG code and by assuming a peak intensity for the $J$=2$\rightarrow$1 transition of 9$\sigma$ (i.e. $\sim$1.8$\,$K; see Table$\,$\ref{tab2}) and a T(3$\rightarrow$2)/T(2$\rightarrow$1) ratio of $\sim$0.4, which corresponds to a H$_2$ density of $\sim$7$\times$10$^3$$\,$cm$^{-3}$, i.e. the average H$_2$ density derived for the filaments.} of N($^{13}$CO)=2$\times$10$^{15}$$\,$cm$^{-2}$, the estimated gas mass missing for Filament 1 is $\sim$300$\,$M$_\odot$ (i.e. $\sim$7\% its total gas mass). After including this gas mass correction, Filament 1 is still 2-3 times less massive than Filaments 2 and 3. 

From Figure$\,$\ref{f10}, we also find that the excitation temperature for the $J$=2$\rightarrow$1 transition, T$_{\rm ex}$(2$\rightarrow$1), ranges from $\sim$7$\,$K to $\sim$14$\,$K in the three filaments. The average values of T$_{\rm ex}$(2$\rightarrow$1) are 10-11$\,$K (Table$\,$\ref{tab5}), showing that the $^{13}$CO gas is sub-thermally excited. These excitation temperatures are similar to those found toward low-mass dark clouds \citep[][]{pin08}. We note however that T$_{\rm ex}$(2$\rightarrow$1) tends to increase to $\sim$13-14$\,$K toward the eastern edge of the IRDC for Filaments 2 and 3, as a consequence of the higher H$_2$ densities found in these regions (Figure$\,$\ref{f10}). This behaviour is similar to that found in \citet{hen13} from N$_2$H$^+$ data. The measured values of T$_{\rm ex}$(3$\rightarrow$2) are typically $\sim$2$\,$K below those reported for T$_{\rm ex}$(2$\rightarrow$1).

The average optical depth obtained from our LVG results for the $^{13}$CO $J$=2$\rightarrow$1 and $J$=3$\rightarrow$2 lines is, respectively, $\tau_{\rm 21}$$\sim$2-3 and $\tau_{\rm 32}$$\sim$1-2 (see Table$\,$\ref{tab5}). This indicates that the $^{13}$CO emission is moderately optically thick. 

\subsection{LVG results for C$^{18}$O}
\label{c18o}

The C$^{18}$O data toward G035.39-00.33 reveal only three positions in Filaments 2 and 3 where the $J$=2$\rightarrow$1 and $J$=3$\rightarrow$2 lines are both brighter than the 5$\sigma$ level in the C$^{18}$O spectra. These positions are: 1) core H6; 2) a C$^{18}$O emission peak toward the west of core H6 (the {\it C$^{18}$O West Peak} at offset (-9,24); Figure$\,$\ref{f2}); and 3) offset (-9,80) located 10$"$ north the region where narrow SiO has been detected \citep[{\it the Narrow SiO Peak}; see][]{jim10}. In Table$\,$\ref{tab6}, we report the H$_2$ density, C$^{18}$O column density, and excitation temperature and optical depth for the $J$=2$\rightarrow$1 and $J$=3$\rightarrow$2 transitions, obtained with the LVG approximation and by using the C$^{18}$O $J$=2$\rightarrow$1 and $J$=3$\rightarrow$2 spectra smoothed within a 14$"$-beam toward each position. From this Table, we find that core H6 is the densest of the three regions with a H$_2$ number density of $\sim$10$^4$$\,$cm$^{-3}$. The H$_2$ densities for the C$^{18}$O West peak and the Narrow SiO Peak are of $\sim$2-4$\times$10$^3$$\,$cm$^{-3}$. The C$^{18}$O column densities are however very similar toward the three positions, with values within a factor of 1.5. The derived excitation temperatures are similar to those measured for $^{13}$CO, and the optical depths are $\leq$1, indicating that the emission from the C$^{18}$O lines is optically thin. This is consistent with the results of \citet{her11}. 

From C$^{18}$O $J$=1$\rightarrow$0 and $J$=2$\rightarrow$1 observations, in \citet{her11} we found that G035.39-00.33 shows widespread CO depletion with depletion factors, $f_D$, as high as $\sim$5. Since the $J$=3$\rightarrow$2 line emission has a higher critical density ($n_{\rm crit}$$\sim$5$\times$10$^4$$\,$cm$^{-3}$) than the $J$=1$\rightarrow$0 and $J$=2$\rightarrow$1 lines ($n_{\rm crit}$$\sim$2$\times$10$^3$$\,$cm$^{-3}$ and $\sim$8$\times$10$^3$$\,$cm$^{-3}$, respectively), our LVG results obtained using the C$^{18}$O $J$=3$\rightarrow$2 data have the potential to provide the depletion factor toward even denser regions with a higher degree of CO depletion. Following the notation used in \citet{her11}, we can calculate the depletion factor, $f_D$, as:

   \begin{equation}
      f_D=\frac{\Sigma_{\rm SMF}}{\Sigma_{\rm C18O}},
   \label{fD}
   \end{equation}
 
\noindent 
where $\Sigma_{\rm SMF}$ denotes the mass surface density derived from near- and mid-IR images \citep{kai13}, and $\Sigma_{\rm C18O}$ is the mass surface density obtained from the observed C$^{18}$O column density toward Filaments 2 and 3. To calculate $\Sigma_{\rm C18O}$, we have used Eq.$\,$5 from \citet{her11}, and assumed an $^{16}$O/$^{18}$O isotopic ratio of 327 \citep{wil94}, and a CO fractional abundance of 2$\times$10$^{-4}$ \citep{lac94}.

In Table$\,$\ref{tab6}, we report the $\Sigma_{\rm SMF}$, the $\Sigma_{\rm C18O}$ and the CO depletion factor measured toward core H6, the C$^{18}$O West Peak, and the Narrow SiO Peak. The derived CO depletion factor is $f_D$$\sim$5-12. In particular, toward core H6, this value is a factor of $\geq$2 higher than those reported in \citet{her11}. The CO depletion factors measured toward G035.39-00.33 are thus similar to those derived toward low-mass pre-stellar cores \citep[e.g.][]{cra05}, and consistent with those recently estimated toward a sample of IRDCs \citep{fon12}. This is expected to largely affect the deuterium chemistry, as found toward massive starless cores where the deuterium fractionation, D$_{\rm frac}$, is increased by a factor of 10 with respect to more evolved high-mass protostellar objects \citep[see][]{fon11}. 

\subsection{Sensitivity of the LVG results to T$_{\rm kin}$}
\label{tkin}

In Sections$\,$\ref{13co} and \ref{c18o}, we have assumed that the kinetic temperature of the gas in G035.39-00.33 is T$_{\rm kin}$=15$\,$K. However, as measured from NH$_3$ observations toward other IRDCs \citep[see e.g.][]{pil06,rag11}, T$_{\rm kin}$ could range from $\sim$11$\,$K to 18$\,$K. We evaluate the effects of T$_{\rm kin}$ on our LVG study by considering a kinetic temperature of $\pm$2$\,$K with respect to T$_{\rm kin}$=15$\,$K. In Table$\,$\ref{tab5}, we compare the LVG results obtained for T$_{\rm kin}$=13$\,$K, 15$\,$K and 17$\,$K. The largest discrepancies between LVG runs are found for the H$_2$ densities, which are higher by up to $\sim$50\% for the case with T$_{\rm kin}$=13$\,$K and lower by up to 25\% for T$_{\rm kin}$=17$\,$K. The derived $^{13}$CO column densities differ by $\sim$4-20\%, while the excitation temperatures vary by less than 5\%. The optical depths only show large discrepancies (by up to a factor of 1.8) for Filament 2 and T$_{\rm kin}$=13$\,$K due to the larger $^{13}$CO column density derived for this filament. In average, the gas mass of the filaments lie within 30\% those calculated for T$_{\rm kin}$=15$\,$K, except for Filament 2 and T$_{\rm kin}$=13$\,$K for which the derived gas mass is a factor of $\sim$2 higher.

\subsection{Effects of including the $J$=1$\rightarrow$0 data in the LVG analysis}
\label{J10}

In this Section, we analyze in what extent our results from Section$\,$\ref{13co} may change by considering the low-excitation $J$=1$\rightarrow$0 line data in the LVG analysis. As examples, we have used the $^{13}$CO line spectra shown in Figure$\,$\ref{f3}, which have been spatially smoothed to the angular resolution of 22$"$ of the $J$=1$\rightarrow$0 data (the Gaussian fits to the lines are reported in Table$\,$\ref{tab3}). We select the offsets (0,20) and (10,-125) as representative of the high-density cores H6 and H3 (Figure$\,$\ref{f2}), and positions (-25,-115) and (0,-50) characteristic of the lower-density regions in the IRDC (hereafter, {\it the lower-density positions}). As before, we assume T$_{\rm kin}$=15$\,$K. 

For the {\it lower-density positions}, our LVG calculations show that the H$_2$ densities and $^{13}$CO column densities derived by using the three $^{13}$CO line transitions are within factors of 2 those obtained in Section$\,$\ref{13co}. However, we find larger discrepancies for the positions toward cores H6 and H3 because the derived H$_2$ densities are factors $\sim$5-20 lower than those estimated from the $J$=2$\rightarrow$1 and $J$=3$\rightarrow$2 lines only. This is expected since the $J$=1$\rightarrow$0 emission probes lower-density gas (with a critical density $n_{\rm crit}$$\sim$2$\times$10$^3$$\,$cm$^{-3}$) than the $J$=2$\rightarrow$1 ($n_{\rm crit}$$\sim$10$^4$$\,$cm$^{-3}$) and $J$=3$\rightarrow$2 transitions ($n_{\rm crit}$$\sim$3$\times$10$^4$$\,$cm$^{-3}$), and therefore arises mainly from the outskirts of the IRDC. The derived $^{13}$CO column densities are larger by up to a factor of $\sim$5 because the amount of material probed by the $J$=1$\rightarrow$0 transition in the direction of the line-of sight is also larger. Therefore, the addition of the $J$=1$\rightarrow$0 line data to the LVG analysis provides good estimates toward regions with relatively low densities of few 10$^3$$\,$cm$^{-3}$. However, it largely underestimates the H$_2$ densities toward the high-density cores (with actual H$_2$ densities $\geq$10$^4$$\,$cm$^{-3}$), as well as it overestimates the total column density of the molecular gas.  

\begin{table*}
 \centering
 \begin{minipage}{150mm}
  \caption{LVG results from C$^{18}$O and CO depletion.}
  \begin{tabular}{lccccccccc}
  \hline
& n(H$_2$) & N(C$^{18}$O) & T$_{\rm ex}$(2$\rightarrow$1) & T$_{\rm ex}$(3$\rightarrow$2) & $\tau_{\rm 21}$ & $\tau_{\rm 32}$ & $\Sigma_{\rm C18O}$ & $\Sigma_{\rm SMF}$ & $f_D$ \\
& (cm$^{-3}$) & (cm$^{-2}$) & (K) & (K) & & & (g$\,$cm$^{-2}$) & (g$\,$cm$^{-2}$) \\ \hline
Core H6 & 1.3$\times$10$^{4}$ & 3.2$\times$10$^{15}$ & 11.7 & 9.8 & 0.4 & 0.3 & 0.025 & 0.30 & $\sim$12 \\ 
C$^{18}$O West Peak & 2.4$\times$10$^{3}$ & 5.0$\times$10$^{15}$ & 7.6 & 6.5 & 0.9 & 0.4 & 0.04 & 0.17 & $\sim$5 \\ 
Narrow SiO Peak & 3.7$\times$10$^{3}$ & 3.0$\times$10$^{15}$ & 8.2 & 7.1 & 0.7 & 0.3 & 0.023 & 0.18 & $\sim$8 \\ \hline
\label{tab6}
\end{tabular}
Note.- $\Sigma_{\rm C18O}$ has been calculated assuming an isotopic ratio $^{16}$O/$^{18}$O of 327 \citep{wil94}, and a CO fractional abundance of 2$\times$10$^{-4}$ \citep{lac94}.
\end{minipage}
\end{table*}

\section{Discussion}
\label{dis}

Since IRDCs represent the initial conditions of massive star and star cluster formation, 
observations of IRDCs allow testing of different theoretical models for (dense) molecular cloud formation. In one scenario of molecular cloud formation, rapid colliding flows of atomic warm gas lead to the formation of cold and filamentary molecular structures due to thermal and dynamical instabilities in the shock front \citep{vaz03,van07,hen08,heit09}. However, another possible scenario is that IRDCs form via GMC-GMC collisions, i.e. of gas that is already mostly molecular \citep{tan00,tas09}. 

From the present data, it is difficult to discriminate between these two scenarios. However, we can provide some constraints on the initial physical conditions and dynamics of the gas in extremely filamentary IRDCs. Our detailed excitation analysis of the $^{13}$CO emission in Section$\,$\ref{13co} shows that the average H$_2$ densities in the filaments of G035.39-00.33 range from $\sim$5000-7000$\,$cm$^{-3}$, where Filaments 2 and 3 are more dense and more massive than Filament 1. The non-thermal velocity dispersion of the $^{13}$CO and C$^{18}$O gas in G035.39-00.33 is supersonic (Section$\,$\ref{width}), similar to what has been found in other IRDCs \citep{rag11,vas11}. This is the observed behaviour in massive molecular clouds and the predicted outcome in both flow-driven \citep[e.g.][]{heit08,car13} and more quiescent models of molecular cloud and massive star/cluster formation \citep[e.g.][]{tan06}. The trend observed for the $^{13}$CO $J$=2$\rightarrow$1 linewidths to get broader toward the outskirts of all filaments in the IRDC reveals dissipation of turbulence and energy at the densest regions in the cloud (Section$\,$\ref{width}).  

One interesting result from the analysis of the individual motions of Filaments 1, 2 and 3 (Section$\,$\ref{grad}), is the detection of a relatively smooth velocity gradient of $\sim$0.4-0.8$\,$km$\,$s$^{-1}$$\,$pc$^{-1}$ in the north-south direction, common to the three molecular filaments (Figure$\,$\ref{f7}). In the following Sections, we discuss several scenarios that could explain the presence of this velocity gradient in Filaments 1, 2 and 3.  

\subsection{Rotation in IRDCs.}
\label{rotation}

\citet{pil06} and \citet{rag12} have also reported the presence of velocity gradients with similar magnitudes toward two samples of IRDCs. For IRDCs with clear signs of active star formation, it is likely that the kinematics of the gas is largely affected by stellar feedback, especially if probed with high-angular resolution interferometric observations \citep[see][]{rag12}. However, this cannot be applied to very filamentary and quiescent IRDCs such as G035.39-00.33 with little star formation. In this Section we explore whether the detected velocity gradient in Filaments 1, 2 and 3 could be produced by the rotation of the cloud. 

The ratio of rotational to gravitational energy, $\beta$, for an elongated cylindrical cloud such as G035.39-00.33 can be estimated as:

   \begin{equation}
      \beta= \frac{E_{\rm rot}}{E_{\rm grav}}\sim\frac{\Omega^2\,L^3}{36GM},
   \label{beta}
   \end{equation}

\noindent
where $E_{\rm grav}$ is the cloud's gravitational energy that can be estimated as $E_{\rm grav}$$\sim$$\frac{3}{2}\frac{GM^2}{L}$ \citep[][]{bas83,bon92}, with $M$ the mass of the cloud, $L$ the length of the filament, and $G$ the gravitational constant. $E_{\rm rot}$ is the kinetic energy due to the cloud's rotation defined as $E_{\rm rot}$=$\frac{1}{2}I\Omega^2$, with $I$ the moment of inertia and $\Omega$ the angular speed of the cloud. For the case of G035.39-00.33, the cloud would rotate around one of the minor axis perpendicular to the main filamentary axis and therefore the moment of inertia of the cylinder is calculated as $I$$\sim$$\frac{ML^2}{12}$ \citep{bon92}.




The velocity gradient measured toward G035.39-00.33 ($\sim$0.4-0.8$\,$km$\,$s$^{-1}$$\,$pc$^{-1}$; Section$\,$\ref{grad}) corresponds to angular speeds of $\Omega_{\rm obs}$$\sim$1-3$\times$10$^{-14}$$\,$s$^{-1}$. By using this value of $\Omega_{\rm obs}$, and from the IRDC's mass of $M$$\sim$2$\times$10$^4$$\,$M$_\odot$ \citep[Section$\,$\ref{13co} and][]{kai13} and the filament's length of $L$$\sim$4.2$\,$pc (Figure$\,$\ref{f1}), we estimate that $\beta$$\sim$0.01. This small value of $\beta$ is consistent with the findings by \citet{rag12}, and indicates that rotation does not dominate the energetics of the IRDC\footnote{We note that the derived value of $\beta$ is also very small ($\beta$$\sim$0.08) if we assume an ellipsoidal cloud for G035.39-00.33 \citep[see also][]{her11b}. For an ellipsoidal cloud, $\beta$ is calculated using Equation$\,$9 in \citet{her11b} and a moment of inertia $I$=$\frac{1}{5}M(r^2+L^2)$, where the cloud's radius is $r$$\sim$0.4$\,$pc, i.e. $\sim$30$"$ as estimated from Figure$\,$\ref{f1}.}. However, this does not rule out that a small rotation could exist in the IRDC causing the relatively smooth velocity gradient seen in G035.39-00.33, while keeping the cloud in virial equilibrium \citep{her12}. This quiescent scenario of cloud rotation however does not explain the global velocity shift of $\sim$0.2$\,$km$\,$s$^{-1}$ measured across G035.39-00.33 between the densest gas in the IRDC and its lower-density envelope \citep[Section$\,$\ref{grad} and][]{hen13}. 



\subsection{Gas accretion flows in G035.39-00.33.}
\label{acc}

An alternative scenario to explain the smooth velocity gradient detected toward G035.39-00.33 would involve global gas accretion along the molecular filaments onto one of the most massive cores in the IRDC, core H6. Core H6 indeed seems to be located at the converging point between Filaments 1, 2 and 3 (Section$\,$\ref{grad}), and it is likely at an early stage of evolution as revealed by the high CO depletion factor measured toward this core ($f_D$$\sim$12; Section$\,$\ref{c18o}). 

In this scenario, the IRDC would be seen with the northern part of the filament situated towards the observer \citep[c.f.][]{kirk13} and with an inclination angle of 45$^\circ$$<$i$<$90$^\circ$ with respect to the line-of-sight. In this configuration, the gas being accreted from the north of core H6 would appear as red-shifted material while the gas toward the south of this core would be blue-shifted. This range of inclination angles is reasonable since the IRDC shows a very filamentary morphology and its inclination must thus be close to the plane of the sky. Gas accretion flows have been reported toward high-mass \citep[][]{bally87,sch10,liu12,per13} and low-mass star forming regions \citep[][]{kirk13} with velocity gradients comparable to those measured in G035.39-00.33 (Section$\,$\ref{grad}).                                        

In the hypothetical scenario that the observed velocity gradient in Filaments 1, 2 and 3 is due to gas accretion along the filaments, an estimate of the gas accretion rate onto core H6 can be provided by using Equation$\,$4 of \citet[][]{kirk13}. Considering a gas mass of $\sim$7500$\,$M$_\odot$ (we consider the mass contained within either the northern or the southern part of Filament 2, i.e. 1/2 the mass of this filament, accreting onto core H6; see Table$\,$\ref{tab5}), an average velocity gradient of $\sim$0.6$\,$km$\,$s$^{-1}$$\,$pc$^{-1}$, and an inclination angle $i$$\sim$45$^\circ$, the derived accretion rate onto core H6 is $\sim$4700$\,$M$_\odot$$\,$Myr$^{-1}$ (or $\sim$5$\times$10$^{-3}$$\,$M$_\odot$$\,$yr$^{-1}$). This accretion rate is of the same order of magnitude as those found in more evolved, high-mass cluster-forming regions \citep[e.g. G20.08-0.14 N or DR21; see][]{gal09,sch10}, although they have been measured at spatial scales $\leq$1$\,$pc. 


We note however that our data do not show evidence for self-absorption in moderately optically thick lines of high-density tracers such as HCO$^+$ or HNC (Section$\,$\ref{infall}), similar to that found by \citet{kirk13} toward Serpens South or by \citet{per13} toward the IRDC SDC335.579-0.272. In addition, the derived mass of core H6 \citep[$\sim$80$\,$M$_\odot$;][]{rath06,but12} represents a small fraction of the total mass of the cloud \citep[$\sim$2$\times$10$^4$$\,$M$_\odot$; Section$\,$\ref{13co} and][]{kai13}, making it unlikely that core H6 governs the global kinematics of the cloud at parsec-scales. In addition, in the global gravitational collapse scenario one would expect to detect a sharper velocity transition closer to core H6 than what is observed, although this may be due to a dilution effect produced by the large beam of our single-dish observations. 

\subsection{Global gravitational collapse: Convergence of Filaments 1, 2 and 3.}
\label{collapse}

A third scenario that could explain the presence of the smooth velocity gradient in G035.39-00.33, would involve the initial formation of sub-structures inside the turbulent molecular cloud (i.e. Filaments 1, 2 and 3), which subsequently converge one onto another as the cloud undergoes global gravitational collapse. This would be in agreement with the scenario proposed in \citet{hen13} where the global velocity red-shift of 0.2$\,$km$\,$s$^{-1}$ measured for the dense molecular gas in this IRDC would be produced by the convergence of Filaments 1 and 3, where Filament 3 is more massive than Filament 1 (Section$\,$\ref{13co}) and thus, it would be more efficient at sweeping-up material in the filament-filament interaction. This interaction would be relatively gentle and the representative velocity of the filament-filament interaction would be the free-fall velocity, $v_{ff}$, defined as:

   \begin{equation}
      v_{\rm ff}= 1.3 \left[m_{\rm l,100}\,\, ln(\frac{r_0}{r})\right]^\frac{1}{2}\sim2\, \rm km\,s^{-1},
   \label{vff}
   \end{equation}

\noindent
where $r_0$ and $r$ are the initial and final radius of the collapsing cloud and m$_{l,100}$ is the mass per unit length in units of 100$\,$M$_\odot$$\,$pc$^{-1}$. For G035.39-00.33, we have considered that m$_{l,100}$$\sim$3 \citep[the mass per unit length is $\sim$300$\,$M$_\odot$$\,$pc$^{-1}$;][]{her12} and $r_0/r$$\sim$2 (i.e. the cloud has experienced contraction so that its size is a factor of 2 smaller at its final condition). Since $v_{\rm ff}$ is comparable to the relative velocities measured between Filaments 1 and 3 ($\sim$2.7-2.9$\,$km$\,$s$^{-1}$; see Table$\,$\ref{tab4}), we conclude that these velocities are consistent with the global gravitational collapse of the cloud. This scenario would also explain why Filaments 1, 2 and 3 show a velocity gradient in the same direction and of similar magnitude (Section$\,$\ref{grad}), since it would be a reminiscence of the initial, undisturbed kinematic structure of the cloud. 

\subsection{Unresolved sub-filament structures in IRDCs}
\label{subfil}

Filamentary structures even smaller than Filaments 1, 2 and 3 may also be present within G035.39-00.33. These structures would resemble the fiber-like morphology recently reported by \citet{hac13} toward the L1495/B213 Taurus region, where the sub-filaments, or {\it fibers}, are separated in velocity space by $\sim$0.5-1.0$\,$km$\,$s$^{-1}$ and have typical lengths of $\sim$0.4$\,$pc. These structures, unresolved in our single-dish observations toward G035.39-00.33, could mimic a velocity gradient similar to that reported from $^{13}$CO and C$^{18}$O (Section$\,$\ref{grad}). This possibility will be addressed by \citet[][]{hen13b} by using high angular resolution interferometric observations of the dense gas toward G035.39-00.33.

\section{Conclusions}
\label{con}

We have carried out a multi-transition analysis of the $^{13}$CO and C$^{18}$O emission toward the very filamentary IRDC G035.39-00.33. We have complemented these data with single-pointing observations of gas infall tracers such as HCO$^{+}$ and HNC to search for infall signatures toward one of the most massive cores in the cloud. The conclusions of our study are: \\

\noindent
i) In agreement with the results from \citet{jim10} and \citet{hen13}, the $^{13}$CO and C$^{18}$O molecular gas in G035.39-00.33 is distributed in three dynamically de-coupled filaments: Filament 1 ($v_{\rm LSR}$$\sim$43.7$\,$km$\,$s$^{-1}$), Filament 2 ($v_{\rm LSR}$$\sim$45.2$\,$km$\,$s$^{-1}$), and Filament 3 ($v_{\rm LSR}$$\sim$46.5$\,$km$\,$s$^{-1}$). \\

\noindent
ii) The massive dense cores reported toward this IRDC \citep{rath06,but12,ngu11} are preferentially found toward the intersecting positions between Filaments 1 and 3, which suggests that these filaments may be interacting. \\

\noindent
iii) A global, relatively smooth velocity gradient of $\sim$0.4-0.8$\,$km$\,$s$^{-1}$$\,$pc$^{-1}$ is measured from north to south in this IRDC. Several possible scenarios, including rotation, global gas accretion along the filaments, global gravitational collapse, and unresolved sub-filamentary structures, are proposed to explain this velocity gradient. \\

\noindent
iv) The linewidths of the molecular gas in the three filaments clearly exceed the thermal contribution, indicating that the gas motions are supersonic ($\sim$2-3 in sonic Mach number). The $^{13}$CO $J$=2$\rightarrow$1 linewidths get broader in the outer regions of the IRDC. This behaviour is similar to that reported toward the low-mass star-forming B5 core \citep{pin10}. \\

\noindent
v) The average H$_2$ densities in the filaments are $\sim$5000-7000$\,$cm$^{-3}$, with Filament 1 being less dense and less massive than Filaments 2 and 3. This would explain the systematic velocity red-shift observed for the high-density gas with respect to the low-density gas in the IRDC envelope reported in \citet{hen13}. \\

\noindent
vi) The CO depletion derived from the high-J C$^{18}$O lines toward G035.39-00.33 are $f_D$$\sim$5-12, factors of 2-3 larger than those previously measured from lower-density C$^{18}$O lines \citep{her11}. \\ 

\section*{Acknowledgments}

We acknowledge the IRAM and JCMT staff, in particular H. Wiesemayer and I. Coulson, for their great help during the observations and reduction of the data. We also thank an anonymous referee for valuable comments that helped to improve the paper. We would like to thank Phil Myers, Qizhou Zhang and Eric Keto for stimulating discussions. The research leading to these results has received funding from the People Programme (Marie Curie Actions) of the European Union's Seventh Framework Programme (FP7/2007-2013) under REA grant agreement number PIIF-GA-2011-301538. P. Caselli acknowledges the financial support of successive rolling grants awarded by the UK Science and Technology Funding Council and of the European Research Council (ERC; project PALs 320620). The work of JK was supported by the Deutsche Forschungsgemeinschaft priority program 1573 ("Physics of the Interstellar Medium"). IRAM is supported by INSU/CNRS (France), MPG (Germany) and IGN (Spain). This paper has made use of JCMT data from Program M08BU35. The James Clerk Maxwell Telescope is operated by the Joint Astronomy Centre on behalf of the Science and Technology Facilities Council of the United Kingdom, the National Research Council of Canada, and (until 31 March 2013) the Netherlands Organisation for Scientific Research.

\appendix

\section[]{Multi-Gaussian profile fits to the $^{13}$CO and C$^{18}$O lines.}
\label{fits}

\begin{figure}
\begin{center}
\includegraphics[angle=0,width=0.45\textwidth]{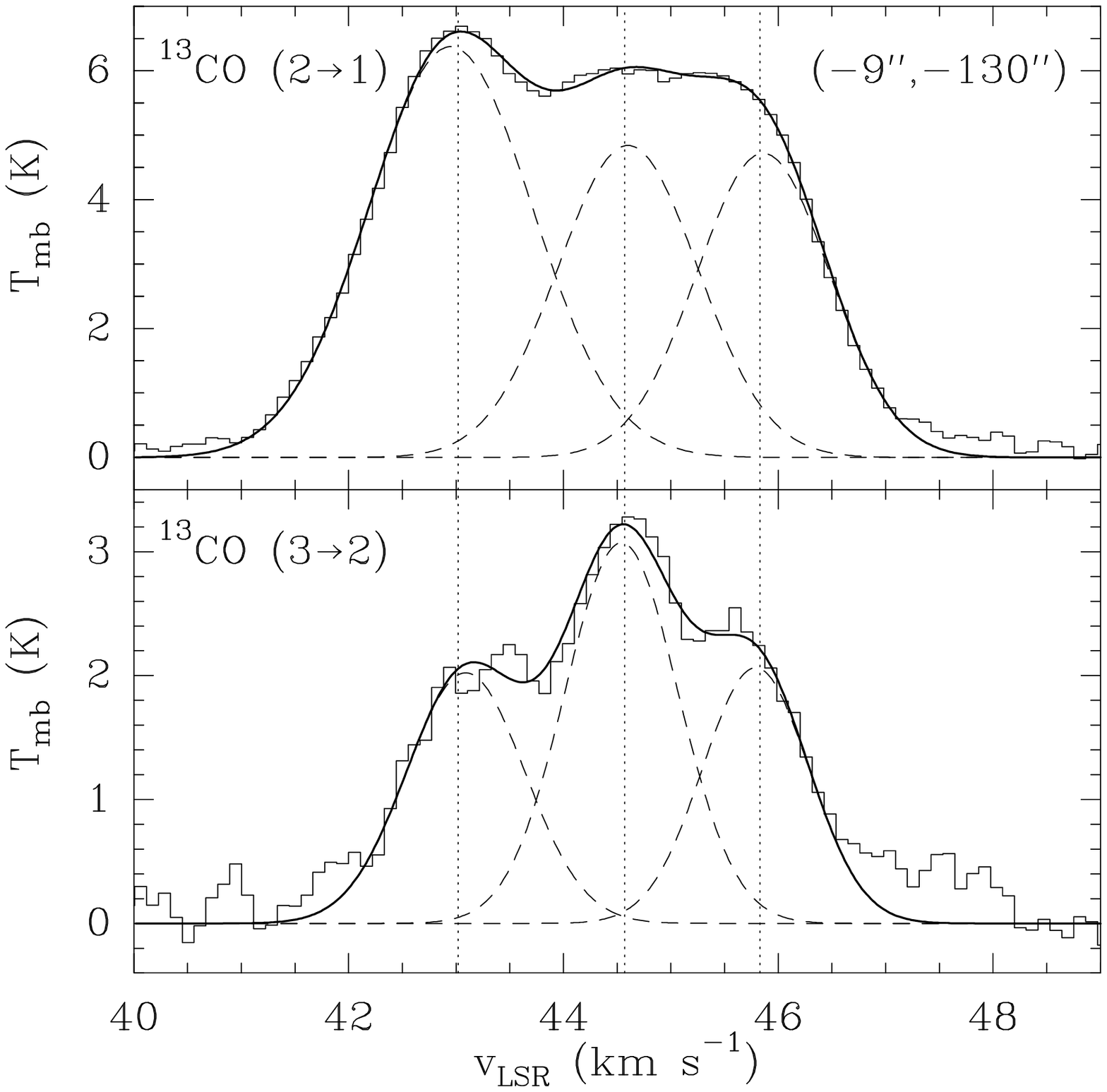}
\caption{Example of the multi-Gaussian profile fitting of the $^{13}$CO $J$=2$\rightarrow$1 and $J$=3$\rightarrow$2 lines. Dashed lines show the individual Gaussian fits of the velocity components associated with Filaments 1, 2 and 3. Thick line shows the total fit obtained by adding the individual Gaussian line profiles from every filament. Vertical dotted lines show the central radial velocity derived for every filament with our Gaussian-fitting method.}
\label{a1}
\end{center}
\end{figure}

The $^{13}$CO and C$^{18}$O line emission toward IRDC G035.39-00.33 shows complex line profiles where three different velocity components are detected along the line-of-sight (Sections$\,$\ref{single} and \ref{chan}). The central radial velocities and linewidths of these components were determined by using a standard multi-Gaussian profile fitting method available within the CLASS software. The line fitting was done by fixing the number of Gaussian components to three, leaving as free parameters the centroid velocities and linewidths of the three Gaussian components. We initially excluded from our analysis the Gaussian fits where the individual velocity components have derived peak intensities with S/N$\leq$3. In order to avoid problems with the uniqueness of the multi-Gaussian fit, especially for the $^{13}$CO lines for which it is more difficult to distinguish between the three velocity components, we have used the following procedure. First, for every position in the map the Gaussian fits of the $J$=2$\rightarrow$1 and $J$=3$\rightarrow$2 spectra were visually inspected one-by-one, and compared one aside the other, to obtain consistent results. In some cases, the Gaussian fitting of the $J$=2$\rightarrow$1 line was aided by that of the $J$=3$\rightarrow$2 data toward the same position (see e.g. Figure$\,$\ref{a1}). Second, after the fitting of the lines, we adopted a detection limit criteria of S/N$\geq$9 for the $^{13}$CO $J$=2$\rightarrow$1 and $J$=3$\rightarrow$2 line components similar to that used by \citet{hen13}. This criteria excludes the data with poor Gaussian fits from our analysis. Note that for C$^{18}$O $J$=2$\rightarrow$1 and $J$=3$\rightarrow$2, the adopted detection limit criteria is S/N$\geq$5 since the line emission is weaker. And third, we compared one-by-one the radial velocities derived from the $J$=2$\rightarrow$1 emission to those obtained from the $J$=3$\rightarrow$2 lines for every velocity component. The data with radial velocities showing discrepancies larger than 0.5$\,$km$\,$s$^{-1}$ (i.e. $\sim$1$/$3 the average linewidth of the $^{13}$CO lines; see Table$\,$\ref{tab4}) were removed from the analysis. The fraction of data excluded at this stage is $\leq$8\%. 

The radial velocities derived for $^{13}$CO and C$^{18}$O were found to fall within the ranges 42.5-44.5$\,$km$\,$s$^{-1}$ for Filament 1, 44.5-46.5$\,$km$\,$s$^{-1}$ for Filament 2, and 45.5-47.5$\,$km$\,$s$^{-1}$ for Filament 3 (see also Table$\,$\ref{tab3}). These velocity ranges are consistent with those considered by \citet{hen13} for the Guided Gaussian Fit Method (see their Appendix). The small differences between the velocity ranges in our method and in their method could be due to the higher angular resolution of our $^{13}$CO and C$^{18}$O data (beam of $\sim$14$"$) compared to their 3$\,$mm data (beam of $\sim$26$"$). We note that similar velocity gradients have been found along G035.39-00.33 for all three filaments by using either the $^{13}$CO or the C$^{18}$O lines (Section$\,$\ref{grad}), which indicates that our Gaussian-fitting method provides robust results.    

\section[]{Intensity-weighted average of $v_{\rm LSR}$ and $\Delta v$.}
\label{aver}

As shown in Section$\,$\ref{grad}, the kinematics of the individual Filaments 1, 2 and 3 in G035.39-00.33 are rather complex with a clear velocity gradient of $\sim$0.4-0.8$\,$km$\,$s$^{-1}$$\,$pc$^{-1}$ in the north-south direction. Despite their complex kinematics, one can derive an average for the radial velocity of the gas in every filament to characterize their global motion. Assuming that the line emission is optically thin, we can then use an intensity-weighted average of $v_{\rm LSR}$ (i.e. $<v_{\rm LSR}>_{\rm F_k}$ with $\rm F_k$ being F$_1$ for Filament 1, F$_2$ for Filament 2, and F$_3$ for Filament 3) that is calculated as:

   \begin{equation}
      <v_{\rm LSR}>_{\rm F_k}=\frac{\Sigma_{\rm i,j}\,\, I_k^{\rm (i,j)}\,\, v_k^{\rm (i,j)}}{\Sigma_{\rm i,j}\,\, I_k^{\rm (i,j)}},
   \label{vave}
   \end{equation}

\noindent
where $v_k^{\rm (i,j)}$ is the central radial velocity derived for Filament $\rm F_k$ toward position $\rm (i,j)$, and $I_k^{\rm (i,j)}$ is the peak intensity of the molecular emission for the same filament and position. By using the peak intensities of the lines as weights, we provide more weight to the values of $v_{\rm LSR}$ derived with a higher accuracy since they correspond to the brightest velocity components detected in the spectra. The average $<v_{\rm LSR}>_{\rm F_k}$ is calculated by considering only those positions where at least two of the velocity components are detected with S$/$N$\geq$9. The average $<v_{\rm LSR}>_{\rm F_k}$ derived for every filament and from every molecular line are shown in Table$\,$\ref{tab4}.

We note that our results for $<v_{\rm LSR}>_{\rm F_k}$ change by less than 0.15\% with respect to those shown in Table$\,$\ref{tab4} if, instead of the peak intensities of the lines, the integrated intensities of the lines (i.e. the areas under the individual Gaussian profiles) are used as weights in Equation$\,$\ref{vave}.

Besides the average values of $v_{\rm LSR}$, we can also calculate the intensity-weighted average of the linewidth of the $^{13}$CO and C$^{18}$O lines across G035.39-00.33, or $<\Delta v>_{\rm Tot}$ (see Figure$\,$\ref{f8} and Section$\,$\ref{width}), to have an idea of the global level of turbulence of the gas in the IRDC. The intensity-weighted average of the linewidth, $<\Delta v>_{\rm Tot}$, is determined as:

   \begin{equation}
      <\Delta v>_{\rm Tot}^{\rm (i,j)}=\frac{I_1^{\rm (i,j)}\,\, \Delta v_1^{\rm (i,j)}+ I_2^{\rm (i,j)}\,\, \Delta v_2^{\rm (i,j)} + I_3^{\rm (i,j)}\,\, \Delta v_3^{\rm (i,j)}}{I_1^{\rm (i,j)}+I_2^{\rm (i,j)}+I_3^{\rm (i,j)}},
   \label{LWtot}
   \end{equation}

\noindent
where $\Delta v_1^{\rm (i,j)}$, $\Delta v_2^{\rm (i,j)}$ and $\Delta v_3^{\rm (i,j)}$ are the measured individual linewidths for the emission from Filaments 1, 2 and 3 toward position $\rm (i,j)$ in the map. In addition, we can obtain an estimate of the individual average linewidth for Filaments 1, 2 and 3 by doing:

   \begin{equation}
      <\Delta v>_{\rm F_k}=\frac{\Sigma_{\rm i,j}\,\, I_k^{\rm (i,j)}\,\, \Delta v_k^{\rm (i,j)}}{\Sigma_{\rm i,j}\,\, I_k^{\rm (i,j)}},
   \label{Dave}
   \end{equation}

\noindent
with $\Delta v_k^{\rm (i,j)}$ the measured linewidth for Filament $k$ toward offset $\rm (i,j)$. The average linewidths derived for every filament from the $^{13}$CO and C$^{18}$O data are given in Table$\,$\ref{tab4}. We note that if the integrated intensities (i.e. the areas under the Gaussian fits) are used in Equations$\,$\ref{LWtot} and \ref{Dave} instead of the peak intensities of the lines, the derived average linewidths are $\sim$4-14\% broader than those shown in Table$\,$\ref{tab4}.

\begin{figure}
\begin{center}
\includegraphics[angle=0,width=0.5\textwidth]{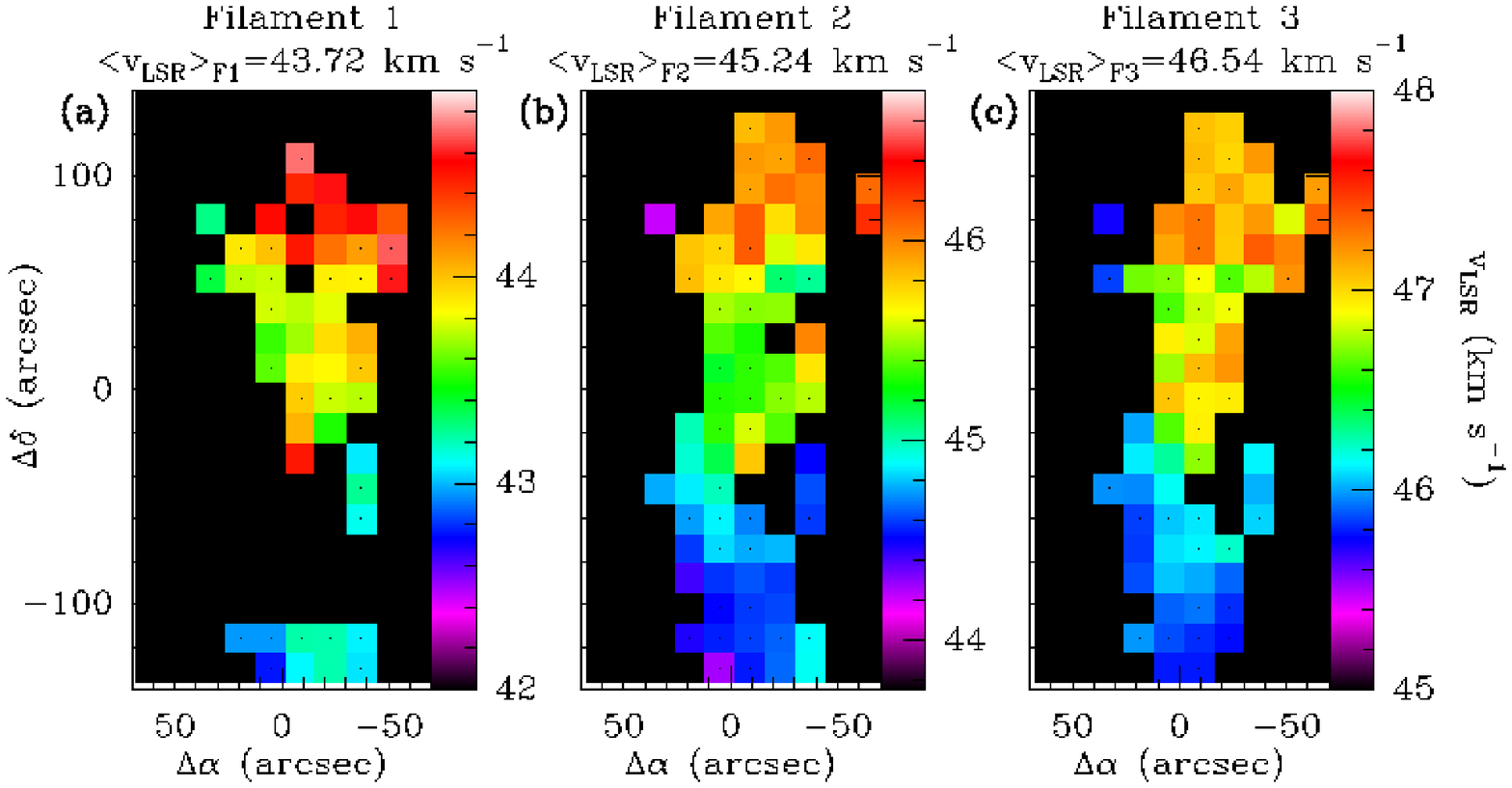}
\caption{As in Figure$\,$\ref{f7}, but obtained from the $^{13}$CO $J$=3$\rightarrow$2 line data. We have only plotted the positions where the derived peak intensity of the $^{13}$CO $J$=3$\rightarrow$2 emission is $\geq$9$\sigma$, with $\sigma$ the rms level in each spectrum.} 
\label{a2}
\end{center}
\end{figure}

\begin{figure}
\begin{center}
\includegraphics[angle=0,width=0.5\textwidth]{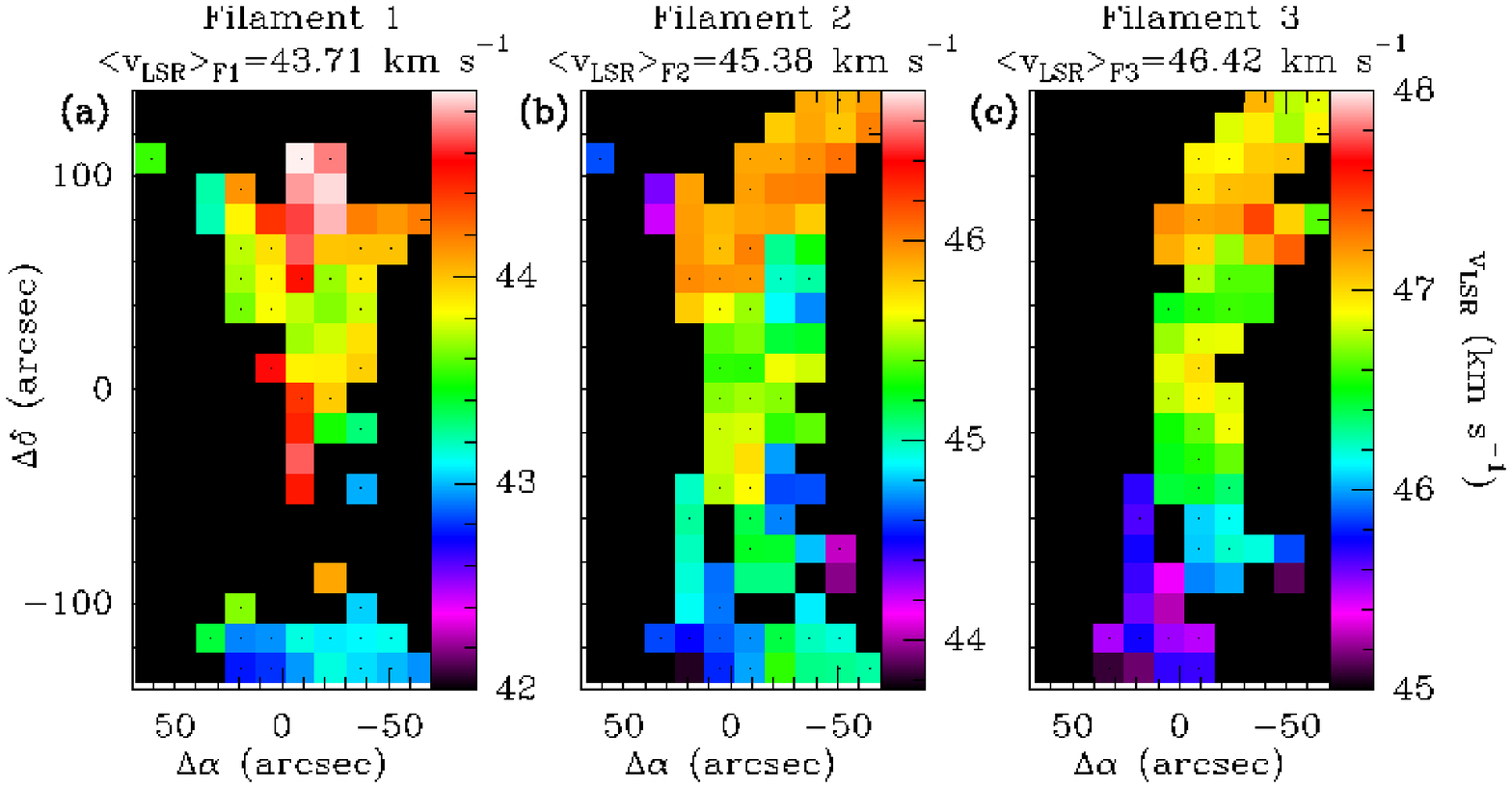}
\caption{As in Figure$\,$\ref{f7}, but obtained from the C$^{18}$O $J$=2$\rightarrow$1 line data. We have only plotted the positions where the derived peak intensity of the C$^{18}$O $J$=2$\rightarrow$1 emission is brighter than 5$\sigma$, with $\sigma$ the rms level in each spectrum.} 
\label{a3}
\end{center}
\end{figure}

\label{lastpage}

\end{document}